\documentclass[reprint,,longbibliography,preprintnumbers,superscriptaddress,prd,
nofootinbib,amsmath,amssymb,aps]{revtex4-1}

\usepackage{graphicx}
\usepackage{dcolumn}
\usepackage{bm}
\usepackage{float}
\usepackage{overpic}
\usepackage[english]{babel}
\usepackage{graphicx}
\usepackage{epsfig}
\usepackage{amssymb}
\usepackage{amsmath}
\usepackage[plainpages=false,pagebackref=false]{hyperref}
\usepackage{ulem}
\usepackage[T1]{fontenc}
\usepackage[utf8]{inputenc}
\usepackage{color}
\usepackage[nameinlink,capitalize]{cleveref}

\usepackage[titletoc]{appendix}
\usepackage{subfig}

\usepackage{booktabs}

\usepackage{comment}

\newcommand{\Tm}{T_{\mathrm{sp}}^-}
\newcommand{\Tp}{T_{\mathrm{sp}}^+}

\begin{document}
\title{Long-lived linear slow roll near critical and spinodal points in soft-wall AdS/QCD}
\author{Han Tang}
\author{Jia-jie Jiang}
\author{Danning Li}
\email{Corresponding author: lidanning@jnu.edu.cn}
\affiliation{ Department of Physics and Siyuan Laboratory, Jinan University, Guangzhou 510632, China}

\begin{abstract}
We systematically investigate the real-time evolution of the chiral condensate near critical and spinodal points within soft-wall AdS/QCD, focusing primarily on the long-lived prethermalization stage, in contrast to the conventional early transients and late-time relaxation. While the quasi-steady nature of this stage has been previously identified, we demonstrate that it is intrinsically a continuous linear slow roll, fundamentally distinct from the asymptotic critical slowing down at late times. By showing that the bulk evolution adiabatically follows the boundary condensate on a slow configuration space, we analytically project the 5D partial differential equation onto a boundary first-order Ginzburg-Landau (GL) equation. We find that the long lifetime of this linear regime is protected by a ``double-zero'' condition: the simultaneous vanishing of the local potential gradient (suppressing the drift velocity) and curvature (suppressing the acceleration). This geometric flatness inherently emerges at the second-order point and both first-order spinodal endpoints. Crucially, this behavior is not a fine-tuned artifact, but a generic consequence of the effective potential landscape intrinsic to these phase-transition endpoints, persisting across a broad class of physically relevant initial conditions and temperature quenches. A local short-time expansion further relates the interval duration to the potential curvature and its higher-order derivatives, in good agreement with our numerical simulations. This work establishes a direct link between the local structure of phase-transition endpoints and the dynamical evolution of the long-lived intermediate stage of the condensate.
\end{abstract}

\maketitle

\section{Introduction}\label{sec:introduction}

Understanding the real-time thermalization of the quark-gluon plasma (QGP) created in relativistic heavy-ion collisions is a key challenge in the study of strongly interacting matter~\cite{Berges:2020fwq}. As the system expands and cools, it probes the region of the QCD phase diagram where chiral symmetry is restored; depending on the collision energy, the system may undergo a crossover, a first-order transition, or pass near a critical point. In any of these scenarios, the chiral condensate---the order parameter associated with this transition region---undergoes a non-equilibrium evolution. Although the condensate itself is not a direct observable, its evolution fundamentally dictates the effective quark masses and the properties of the medium, leaving distinctive imprints on final-state hadron yields, fluctuation observables~\cite{Sakai:2025xmo}, and electromagnetic probes~\cite{Lee:2000ym}. Consequently, establishing the dynamical laws that govern the non-equilibrium evolution of the chiral order parameter across the phase transition is essential for fully bridging theoretical calculations and experimental measurements.

A substantial body of work has addressed the real-time evolution of the chiral condensate within effective models, treating the order parameter as a dynamical variable coupled to a thermal or expanding medium. The framework of nonequilibrium chiral fluid dynamics provides a systematic approach to the coupled evolution of the order parameter and a quark fluid~\cite{Paech:2003fe}. Using the two-particle irreducible effective action, one derives a Langevin equation for the \(\sigma\) field with a damping coefficient and a noise correlator determined by the interaction with the heat bath~\cite{Nahrgang:2011mg}. Within this QCD framework, it has been established that a first-order transition exhibits phase coexistence and substantial supercooling, while at a critical point the relaxation time becomes long~\cite{Nahrgang:2011mv}, a manifestation of the general phenomenon of critical slowing down in dynamic critical phenomena~\cite{Hohenberg:1977ym}. Importantly, this relaxational process is accompanied by the reheating of the medium, an effect that has been shown to alter the dynamical pathway away from the equilibrium free-energy landscape~\cite{Nahrgang:2011mv,Herold:2018ptm}. The inclusion of a Polyakov loop on the same footing further extends this description to simultaneously propagate the confinement and chiral order parameters~\cite{Herold:2013bi}.

When the order parameter is allowed to evolve in space as well as time, the dynamics becomes much richer. Time-dependent Ginzburg--Landau studies based on the nonlocal Nambu--Jona-Lasinio model show that the system can pass through long-lived, spatially modulated metastable structures, with lifetimes comparable to the time scales of a heavy-ion collision~\cite{Carlomagno:2018ogx}. Spatial inhomogeneity also leaves distinct imprints on the macroscopic flow: when the condensate is coupled to an expanding quark system through the Vlasov equation, the chiral phase transition generates a characteristic kink in the flow velocity and a spatial phase boundary that acts as a ``wall'' for quarks~\cite{Wang:2020wwm}. External conditions can further reshape the dynamics: under a strong magnetic field, the closed-time-path formalism produces a mean-field Langevin equation in which the magnetic field increases the damping coefficient, enhancing fluctuations and stretching the short-time evolution~\cite{Krein:2021sco}.

A complementary line of inquiry focuses on quench dynamics and the disoriented chiral condensate (DCC). The concept of DCC, where the chiral field is misaligned from its true vacuum, was developed through the notion of classical pion fields~\cite{Anselm:1989pk,Anselm:1991pi,Bjorken:1992,Bjorken:1992dcc,Blaizot:1992at}. A pivotal development came from Refs.~\cite{Rajagopal:1992qz,Rajagopal:1993ah}, which first demonstrated that a rapid quench can parametrically amplify long-wavelength pion modes, providing the foundational dynamical mechanism for DCC formation. Subsequent self-consistent treatments with quantum fluctuations and backreaction showed that the growth is limited, and that energy is transferred to fluctuation modes, leading to relaxation of the condensate~\cite{Boyanovsky:1994yk,Cooper:1994ji}. More recent simulations of the $O(4)$ Model G reveal a richer three-stage structure after a quench---initial exponential growth, parametric enhancement, and slow relaxation---linked to an enhanced soft pion yield~\cite{Florio:2025lvu,Florio:2025zqv}. From a systematic perspective, the real-time functional renormalization group has been formulated for the stochastic hydrodynamic equations of Model G (the dynamic universality class of the chiral transition in the two-flavor chiral limit) and Model H (the QCD critical point), deriving nonperturbative RG flow equations for the kinetic coefficients and revealing the fixed-point structure and dynamic critical exponents of these universality classes~\cite{Roth:2024rbi,Roth:2024hcu}.

Beyond the early-stage parametric amplification and critical relaxation, it is worth noting that in 4D weakly coupled quantum field theories, the post-quench evolution often features a prolonged intermediate stage. This prethermal stage is governed by non-thermal fixed points---specifically the quantum and classical prethermal fixed points discussed in Refs.~\cite{Chiocchetta2015,Chiocchetta2017}. However, bridging this weak-coupling prethermalization paradigm to the strongly coupled regime remains an open challenge.

The above effective descriptions establish that the real-time evolution of the chiral condensate is a multi-stage, multi-scale process: it exhibits critical slowing down near the critical point, phase coexistence and supercooling at a first-order transition, spatially modulated metastable structures when inhomogeneity is included, and a characteristic sequence of growth, parametric enhancement, and slow relaxation after a quench. These effective descriptions successfully map out the rich macroscopic phenomenology of the transition dynamics, including their non-perturbative aspects as captured by functional renormalization group methods. A complementary perspective is offered by holographic duality, which provides a geometric framework where the equilibrium phase structure and the real-time non-equilibrium dynamics are not introduced independently, but are both derived from the same higher-dimensional gravitational action.

The AdS/CFT correspondence maps strongly coupled gauge theories onto classical gravity~\cite{Maldacena:1997re,Gubser:1998bc,Witten:1998qj}. A landmark result is the shear-viscosity-to-entropy ratio of the strongly coupled $\mathcal{N}=4$ SYM plasma, \(\eta/s = 1/(4\pi)\)~\cite{Policastro:2001yc,Buchel:2003tz,Kovtun:2004de},
which established that holographic plasmas are naturally close to the perfect-fluid limit indicated by heavy-ion data. A particularly profound insight from nonequilibrium holography is the emergence of a holographic attractor in real-time dynamics. In strongly coupled plasmas, the causal event horizon acts as an ideal absorber for high-frequency excitations, causing the system to rapidly lose memory of its initial conditions and dynamically collapse onto a lower-dimensional invariant manifold---often referred to as the hydrodynamic attractor~\cite{Heller:2011ju,Heller:2013fn}. This raises a natural geometric question: near a chiral phase transition, how does this attractor behavior manifest for the order parameter, and is it dictated by the local phase structure? Within this geometric picture, once the bulk action and geometry are fixed, the static phase diagram and the real-time evolution (e.g., the damping and driving forces) are determined simultaneously. This self-consistency makes holography a particularly useful framework for investigating how the local structure of a phase-transition endpoint dictates the real-time slow dynamics.

To address chiral symmetry breaking within this framework, bottom-up AdS/QCD models provide a suitable starting point. These models have achieved considerable phenomenological success in describing the hadron spectrum, reproducing the masses of light mesons and the linear Regge trajectories (see, e.g., Refs.~\cite{Erlich:2005qh,Karch:2006pv,Brodsky:2006,Gherghetta:2009,Kelley:2011,Li:2013,Li:2013oda,Sui:2010,Sui:2011,Colangelo:2008us,Ballon-Bayona:2023,FolcoCapossoli:2020imm,Capossoli:2016glueball,Gutsche:2013,Rinaldi:2018,Rinaldi:2022}). For instance, the hard-wall model of Ref.~\cite{Erlich:2005qh} realized spontaneous chiral symmetry breaking through a bulk scalar field whose boundary value acts as the quark mass and whose normalizable mode gives the chiral condensate. The soft-wall model (often referred to as the KKSS model) retained the conformal AdS geometry but introduced a quadratic dilaton profile to reproduce linear Regge trajectories~\cite{Karch:2006pv}. The soft-wall construction is particularly convenient for dynamical studies: the chiral condensate is automatically encoded in the near-boundary expansion of the bulk scalar, and the model parameters are constrained by the meson spectrum and equilibrium thermodynamics, so that a single set of parameters governs both the static phase structure and the real-time evolution. Beyond the hadron spectrum, these models have also been successfully employed to map out the equilibrium chiral phase diagram, capturing the crossover, first-order transition, and critical point at finite temperature and density~\cite{Chelabi:2016jhep,Chelabi:2016prd,Li:2017,Bartz:2016,Bartz:2018,Fang:2018,Chen:2019rez,Fang:2016,Colangelo2012Density}. This equilibrium phase structure provides a natural starting point for studying the real-time dynamics of the chiral transition within the same framework.

Real-time chiral dynamics in holographic models has been explored along several lines. Early studies focused on the first-order transition. Ref.~\cite{Evans:2010xs} studied a D3/D7 system with a magnetic field in an expanding or contracting plasma background, going beyond the adiabatic condensate trajectory to solve the full time-dependent partial differential equations for the non-equilibrium transition. Ref.~\cite{Guralnik:2011xz} numerically simulated a first-order chiral phase transition in an expanding, cooling strongly coupled plasma within Lorentzian AdS/CFT, establishing a connection between the latent heat of the transition and the dual gravitational energy. More recently, attention has shifted to the soft-wall model and the prethermalization phenomenon. Ref.~\cite{Cao:2023} quenched the two-flavor soft-wall model and solved the nonequilibrium evolution of the chiral condensate, identifying exponential relaxation away from $T_c$, critical slowing down at $T_c$ with a dynamic critical exponent $z\approx2$, and---for a broad class of initial states---an extended quasi-steady prethermalization stage. Ref.~\cite{Giannuzzi:2025} studied the condensate in a time-dependent black-hole background, showing that prethermalization emerges in the chiral limit when the initial state is prepared near the critical temperature. Ref.~\cite{Zheng:2025} extended the analysis by coupling the condensate to Goldstone modes, finding that an additional prethermalization stage emerges at noncritical temperatures. Ref.~\cite{Tang:2025} extended the real-time study to a three-flavor ($N_f=2+1$) soft-wall model with a Columbia-plot-like phase structure, reporting nontrivial thermalization behavior at intermediate times near the phase transition.

These holographic studies have firmly established that the soft-wall model captures both the relaxation dynamics and the prethermalization phenomenon. However, the fine-grained internal dynamical structure of this intermediate stage---whether it is a strictly static plateau or a continuous drift---has remained unresolved. In particular, it remains unclear why the prethermalization stage persists for such extended time scales, and how its longevity connects to the local geometry of the phase-transition endpoints. While holographic studies of superfluids and critical models have explored near-critical Ginzburg-Landau (GL) descriptions via adiabatic expansions or
numerical matching \cite{Basu:2011ft, Bu2021GinzburgLandau, Donos:2022xfd, Flory:2024}, a systematic analytic projection of the 5D real-time bulk dynamics onto a boundary GL equation is still lacking for the chiral transition in soft-wall AdS/QCD, particularly in extending the analysis beyond symmetric second-order points to first-order spinodal endpoints.

Therefore, in this work, we systematically analyze the real-time initial-value problem in the soft-wall AdS/QCD model, with a focus on the intermediate prethermalization stage. By showing that the bulk evolution is dynamically slaved to a slow configuration space (analogous to the hydrodynamic attractor), we analytically project the 5D real-time bulk dynamics onto an
effective boundary first-order Ginzburg-Landau (GL) equation. This allows us to derive the effective potential directly from the model geometry, rather than treating it as a phenomenological input. Crucially, we reveal that the previously
observed quasi-steady plateau is intrinsically a continuous linear slow roll---fundamentally distinct from the asymptotic critical slowing down at late times---and we elucidate that its long lifetime is geometrically protected by the ``double-zero'' condition at the phase-transition endpoints.

The paper is organized as follows. In Sec.~\ref{sec:model}, we introduce the soft-wall AdS/QCD model and map out the equilibrium phase diagram in the chiral limit. In Sec.~\ref{sec:second_dynamics}, we formulate the real-time
initial-value problem, derive the effective Ginzburg-Landau equation via the slow configuration space, and show how the ``double-zero'' condition yields a long-lived linear slow roll at \(T_c\). In Sec.~\ref{sec:first_order_dynamics},
we extend the analysis to the first-order transition, demonstrating that long-lived linear slow rolls robustly emerge at both spinodal endpoints, with their drift directions dictated by the distinct local geometries. In Sec.~\ref{sec:landscape}, we construct the effective potential landscape and analytically derive the scaling of the slow-roll duration. Finally, we summarize our results in Sec.~\ref{sec:conclusions}. Additional numerical details and supplementary technical derivations are provided in the appendices.

\section{The Model and Equilibrium Phase Structure}\label{sec:model}
Understanding the real-time dynamics of the chiral phase transition in strongly coupled QCD requires a non-perturbative framework. The AdS/CFT correspondence, specifically the bottom-up soft-wall AdS/QCD model developed by Karch, Katz, Son, and Stephanov (KKSS)~\cite{Karch:2006pv}, provides an ideal setting for this purpose. In this section, we briefly review this model and establish the setup used for our equilibrium calculations.

\subsection{Model setup and equation of motion at equilibrium}

The KKSS model reproduces the linear Regge trajectories of light mesons and successfully captures spontaneous chiral symmetry breaking. The original model is based on an \(SU(2)_L \times SU(2)_R\) gauge symmetry in the bulk, where the chiral symmetry breaking is encoded in a \(2\times 2\) matrix-valued scalar field \(X\), transforming as a bifundamental representation. The gauge fields are the left- and right-handed \(SU(2)\) fields \(L_M\) and \(R_M\). To introduce a non-zero baryon chemical potential, we extend the gauge group to \(U(1)_B \times SU(2)_L \times SU(2)_R\), where the \(U(1)_B\) vector field \(V_M\) provides the chemical potential.

Since our primary focus is the homogeneous chiral order parameter and we do not consider other quantum numbers such as isospin, the \(SU(2)\) gauge fields \(L_M\) and \(R_M\) are not turned on as background fields. The relevant dynamical field is therefore the matrix-valued scalar field \(X\), dual to the bilinear operator \(\bar{q}^\alpha q^\beta\) (with \(\alpha, \beta\) the flavor indices). The action for the scalar sector is
\begin{equation}
\label{eq:flavor_action}
S_X=-\int d^5x\sqrt{-g}\,e^{-\Phi} \operatorname{Tr}\left[(D_MX)^\dagger D^M X+V_X(X, F^2)\right],
\end{equation}
where \(D_M\) is the standard covariant derivative following the KKSS construction~\cite{Karch:2006pv}, \(g\) is the determinant of the metric, \(\Phi\) is the dilaton field, and \(V_X(X, F^2)\) is the scalar potential including its interaction with the gauge field strength \(F_{MN} = \partial_M V_N - \partial_N V_M\).

For the background geometry, we consider a fixed Reissner-Nordstr\"om AdS$_5$ black-hole background (with the AdS radius set to unity):
\begin{equation}
 ds^2=e^{2A_s(z)}\left[-f(z)\,dt^2+\frac{dz^2}{f(z)}+d\bm{x}^2\right], \quad A_s(z)=-\ln z .
\end{equation}
Here, the radial coordinate \(z\) ranges from \(z=0\) (the UV conformal boundary) to \(z=z_h\) (the event horizon), where \(f(z_h)=0\). To introduce a finite baryon chemical potential, we turn on the temporal component of the \(U(1)_B\) gauge field:
\begin{equation}
 V_M = A_t(z)\delta_M^t = \mu\left(1-\frac{z^2}{z_h^2}\right)\delta_M^t,
\end{equation}
where \(\mu\) denotes the baryon chemical potential. The blackening factor \(f(z)\) and the temperature \(T\) are given by
\begin{align}
 f(z)&=1-(1+\mu^2z_h^2)\frac{z^4}{z_h^4}+\mu^2z_h^2\frac{z^6}{z_h^6}, \label{eq:f_z}\\
 T&=\frac{1}{\pi z_h}\left(1-\frac{\mu^2z_h^2}{2}\right). \label{eq:T}
\end{align}
The dilaton profile \(\Phi(z)\) is chosen to be
\begin{equation}
 \Phi(z)=-\mu_1z^2+\mu_2z^2\tanh(\mu_3z^2),
\end{equation}
which interpolates between the required UV conformal behavior and the IR linear confinement~\cite{Chelabi:2016jhep,Chelabi:2016prd}.

Working in the degenerate two-flavor case at finite temperature and density, we can reduce the matrix field \(X\) to its diagonal component \(X = \frac{\chi}{2}\mathbf{I}_2\), where \(\chi\) is a real scalar field. The action then simplifies to
\begin{equation}
\label{eq:action}
S_\chi=-\int d^5x\sqrt{-g}\,e^{-\Phi}
\left[\frac12g^{MN}\partial_M\chi\partial_N\chi+U(\chi)\right],
\end{equation}
with the effective potential
\begin{equation}
\label{eq:bulk_potential}
U(\chi)=v_2(z)\chi^2+v_4(z)\chi^4+v_6(z)\chi^6.
\end{equation}
Here, for notational brevity, we use \(U(\chi)\) to denote the effective potential, keeping in mind that its coefficients \(v_i(z)\) explicitly depend on the radial coordinate \(z\) through the background fields. Following Refs.~\cite{Chen:2019rez,Chen2022Rotation}, to reproduce the correct \(T-\mu\) phase diagram, we introduce gauge-invariant non-minimal couplings in the scalar potential and absorb the background medium effects into the coefficients:
\begin{align}
v_2(z)&=-\frac32+\alpha\frac{z^2A_t^2(z)}{f(z)}
+\lambda_2 z^4[A_t'(z)]^2,\label{eq:v2}\\
v_4(z)&=v_{40}+\lambda_4z^4[A_t'(z)]^2,\label{eq:v4}\\
v_6(z)&=v_{60}+\lambda_6z^4[A_t'(z)]^2.\label{eq:v6}
\end{align}
The physical origins of these coefficients are as follows. The constant term \(-3/2\) comes from the standard 5D mass prescription for the dual scalar operator (\(m_5^2=-3\)). The constant coefficients \(v_{40}, v_{60}\), together with the density-dependent couplings \(\lambda_2,\lambda_4, \lambda_6\), arise from the non-minimal field-strength interactions \(F_{MN}F^{MN}\operatorname{Tr}(X X^\dagger)^n\). The term with coefficient \(\alpha\), proportional to \(A_t^2(z)/f(z)\), represents the medium-induced effective mass correction for the condensate \cite{Chen:2019rez,Chen2022Rotation}.

Varying the action \eqref{eq:action} with respect to \(\chi\), we obtain the static equation of motion:
\begin{equation}\label{eq:eom_static}
 \chi''+\left(3A_s'-\Phi'+\frac{f'}f\right)\chi'
       -\frac{e^{2A_s}}f\partial_\chi U(\chi)=0,
\end{equation}
where primes denote derivatives with respect to \(z\). This second-order ordinary differential equation forms a boundary-value problem. 

Since our primary goal is to establish the qualitative theoretical framework and elucidate the mechanisms underlying the phase transition, we work strictly in the chiral limit (\(m_q=0\)) throughout this work. This limit preserves the exact \(Z_2\) chiral symmetry under \(\chi \to -\chi\), allowing for a clean identification of the critical points. A detailed quantitative analysis incorporating physical quark masses, which explicitly breaks this symmetry, is deferred to future work. Near the UV boundary \(z \rightarrow 0\), the scalar field admits the asymptotic expansion
\begin{equation}\label{eq:uv_short}
\chi(z)=m_q \zeta z+\frac{\sigma}{\zeta}z^3+\cdots,
\end{equation}
where \(\sigma\) is the chiral condensate, and \(\zeta=\sqrt{3}/(2\pi)\) is a normalization constant~\cite{Cherman:2008eh}. Because the non-normalizable source term \(m_q \zeta z\) vanishes identically in the chiral limit, the normalizable coefficient \(\sigma\) acts as the physical order parameter. At the IR horizon \(z=z_h\), the field must satisfy the regularity condition, which we implement by expanding \(\chi(z)=c_0+c_1(z_h-z)+\cdots\) and requiring the series to be smooth.

While the model parameters can in principle be constrained by the meson spectrum and equilibrium thermodynamics, our focus here is on the qualitative structure of the phase transition and its real-time dynamics. Therefore, we adopt a representative parameter set that captures the correct phase structure and gives reasonable transition points. The specific values are summarized in Table~\ref{tab:parameters}.

\begin{table}[tbp]
\caption{Model parameters. The three dilaton coefficients have units of \(\mathrm{GeV}^2\); the other entries are dimensionless in the stated normalization.}\label{tab:parameters}
\centering
\begin{ruledtabular}
\begin{tabular}{ll}
\(\mu_1=0.6889\,\mathrm{GeV}^2\) & \(\mu_2=0.8738\,\mathrm{GeV}^2\)\\
\(\mu_3=(1849/60000)\,\mathrm{GeV}^2\) & \(\alpha=-0.1\)\\
\(\lambda_2=-0.1\) & \(\zeta=\sqrt{3}/(2\pi)\)\\
\(v_{40}=10\) & \(\lambda_4=-15\)\\
\(v_{60}=100\) & \(\lambda_6=300\)
\end{tabular}
\end{ruledtabular}
\end{table}

Due to the nonlinearities in \(U(\chi)\), Eq.~\eqref{eq:eom_static} generally admits multiple branches of solutions. We solve this boundary-value problem numerically using the standard shooting method. We initialize the numerical integration from both the UV boundary (with a given \(\sigma\)) and the IR horizon (with a given \(c_0\)), and match the field and its first derivative at an intermediate point (\(z=0.5z_h\)). By scanning the parameter space, we resolve the stable, metastable, and unstable branches, which lay the groundwork for constructing the equilibrium phase diagram.

\subsection{Equilibrium phase diagram in the chiral limit}

To establish the thermodynamic context for the subsequent real-time dynamics, we now map out the equilibrium phase diagram in the \(T-\mu\) plane by solving the static equation of motion \eqref{eq:eom_static}.

\begin{figure}[tbp]
\centering
\includegraphics[width=0.98\columnwidth]{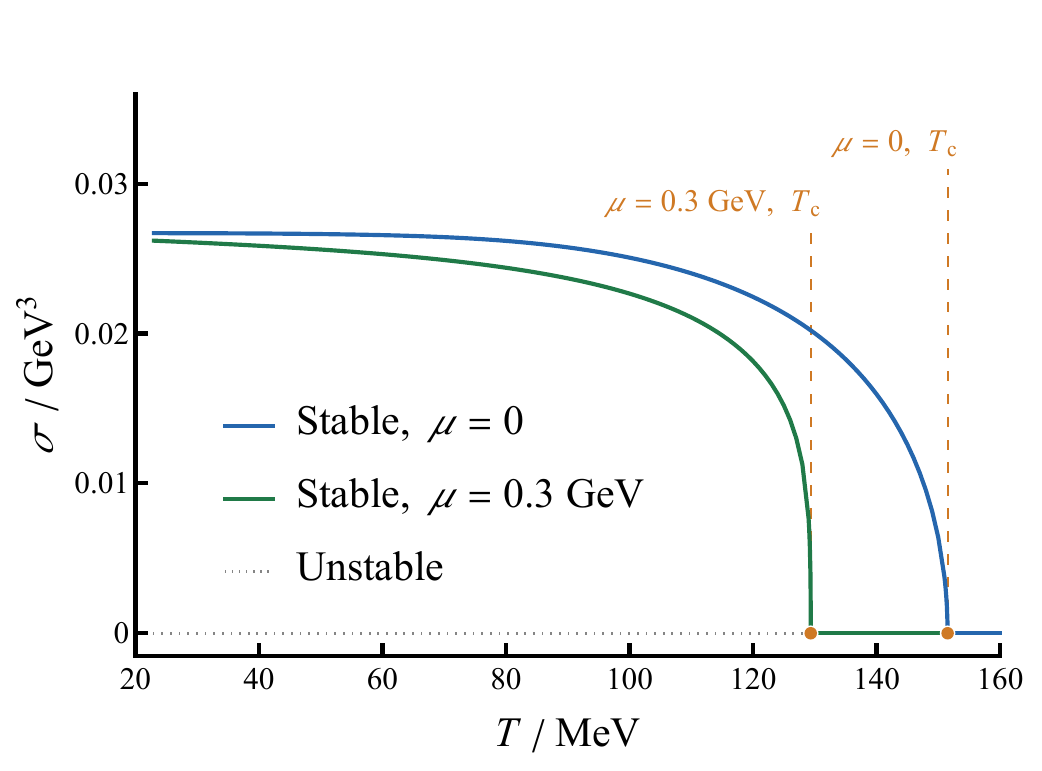}
\caption{Static chiral condensate for the second-order transitions. The blue curves denote the case at \(\mu=0\), and the green curves denote a small chemical potential \(\mu = 0.3\,\mathrm{GeV}\). Solid and dotted curves denote stable and unstable branches, respectively. The marked temperatures identify the corresponding critical points \(T_c\).}
\label{fig:equilibrium_second}
\end{figure}

At \(\mu=0\), the static condensate vanishes continuously at \(T_c(\mu=0) \simeq 0.151515645\,\mathrm{GeV}\). The nonzero branch is stable below \(T_c\), whereas the zero-condensate branch is unstable below \(T_c\) and stable above it. When a small chemical potential is turned on (e.g., \(\mu = 0.3\,\mathrm{GeV}\)), the system remains in the second-order regime, but the critical temperature decreases. As shown in Fig.~\ref{fig:equilibrium_second}, the condensate still vanishes continuously at a shifted critical temperature \(T_c(\mu=0.3\,\mathrm{GeV}) \simeq 0.129353781\,\mathrm{GeV}\), demonstrating that the continuous nature of the transition is preserved for \(\mu < \mu_{\rm TCP}\simeq 0.304\,\mathrm{GeV}\). These stable and unstable behaviors are depicted by the solid and dotted curves in Fig.~\ref{fig:equilibrium_second}, respectively.

\begin{figure}[tbp]
\centering
\includegraphics[width=0.98\columnwidth]{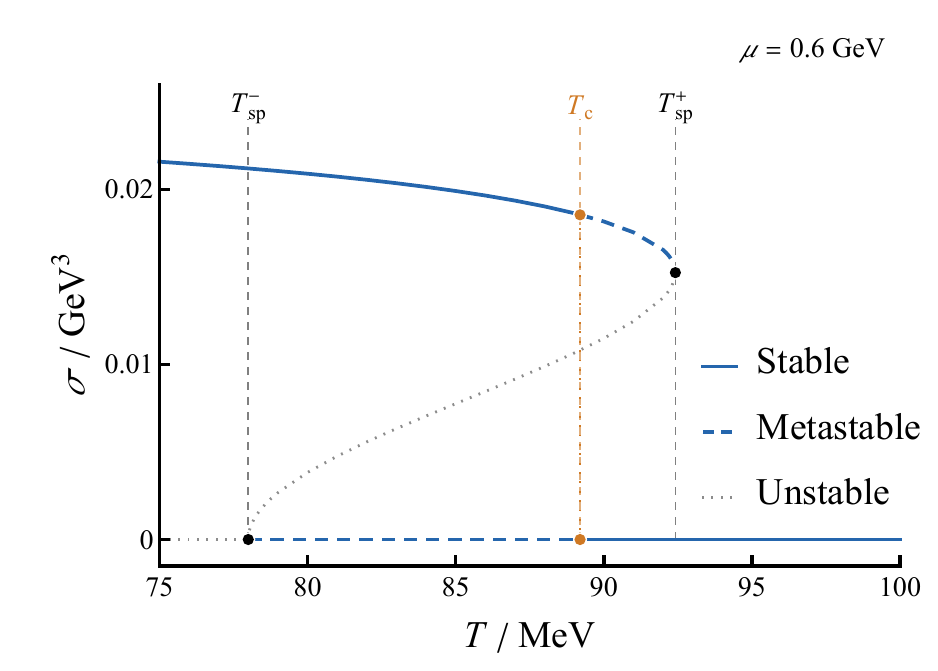}
\caption{Static chiral condensate branches at \(\mu=0.6\,\mathrm{GeV}\). Solid, dashed, and dotted curves denote stable, metastable, and unstable branches, respectively. The marked temperatures identify coexistence at \(T_c\) and the two spinodal endpoints \(T_{\mathrm{sp}}^-\) and \(T_{\mathrm{sp}}^+\).}
\label{fig:first_order_static}
\end{figure}

As the chemical potential increases further and exceeds the tricritical value \(\mu_{\rm TCP} \simeq 0.304\,\mathrm{GeV}\), the nature of the transition changes. At \(\mu=0.6\,\mathrm{GeV}\), the equilibrium chiral condensate undergoes a first-order transition. Figure~\ref{fig:first_order_static} shows the static branches. The unstable branch meets the zero-condensate branch at the lower spinodal \(\Tm\) and the nonzero branch at the upper spinodal \(\Tp\). Specifically, at this chemical potential, we obtain \(\Tm \simeq 0.0780005935\,\mathrm{GeV}\) and \(\Tp \simeq 0.0924180303\,\mathrm{GeV}\). Within the multivalued region bounded by \(\Tm\) and \(\Tp\), the static equation admits multiple solutions (stable, metastable, and unstable), signaling a change in the transition nature.

To determine which of these multiple branches is thermodynamically stable, we compute the grand potential of the system. Following the standard holographic prescription, the on-shell Euclidean action relates to the thermodynamic potential. At fixed temperature and chemical potential, the scalar contribution to the grand potential per unit spatial volume is given by the regulated expression
\begin{equation}\label{eq:free_energy_def}
\begin{split}
 \mathcal{F}_\varepsilon
 &=\frac{T S_{E,\varepsilon}}{V_3}\\
 &=\int_\varepsilon^{z_h}dz\,e^{5A_s-\Phi}
 \left[\frac12 e^{-2A_s}f(\partial_z\chi)^2+U(\chi)\right],
\end{split}
\end{equation}
where \(S_{E,\varepsilon}\) is the regularized Euclidean action and \(V_3\) is the spatial volume.

Integrating by parts and using the equation of motion \eqref{eq:eom_static}, the on-shell grand potential can be expressed as
\begin{equation}\label{eq:free_energy_onshell}
\begin{split}
 \mathcal{F}_\varepsilon={}&\int_\varepsilon^{z_h}dz\,e^{5A_s-\Phi}
 \left(U(\chi)-\frac12\chi\partial_\chi U(\chi)\right)\\
 &-\left.\frac12 e^{3A_s-\Phi}f\chi\partial_z\chi\right|_{z=\varepsilon}.
\end{split}
\end{equation}
In the chiral limit, this expression is finite as the UV cutoff \(\varepsilon\to0\). The thermodynamically stable state is the one that minimizes \(\mathcal{F}\). When the grand potentials of two distinct branches become equal, \(\mathcal{F}_{\text{branch1}} = \mathcal{F}_{\text{branch2}}\), the system undergoes a first-order coexistence at temperature \(T_c\). As a representative example, at \(\mu=0.6\,\mathrm{GeV}\), this condition yields \(T_c(\mu=0.6\,\mathrm{GeV}) \simeq 0.0892003801\,\mathrm{GeV}\), which indeed lies between the two spinodal temperatures \(\Tm\) and \(\Tp\).

Within the multivalued region, the local stability of each branch is determined by the curvature of the effective potential. The boundaries of this region are defined by the spinodal temperatures. At the lower spinodal \(\Tm\), the zero-condensate state loses stability. At the upper spinodal \(\Tp\), the nonzero-condensate state loses stability. Consequently, within the temperature interval bounded by \(\Tm\) and \(\Tp\), the solid and dashed curves represent the globally stable and locally stable metastable states, respectively, while the unstable branch connecting the two spinodal endpoints is denoted by the dotted curve in Fig.~\ref{fig:first_order_static}.

\begin{figure}[tbp]
\centering
\includegraphics[width=\linewidth]{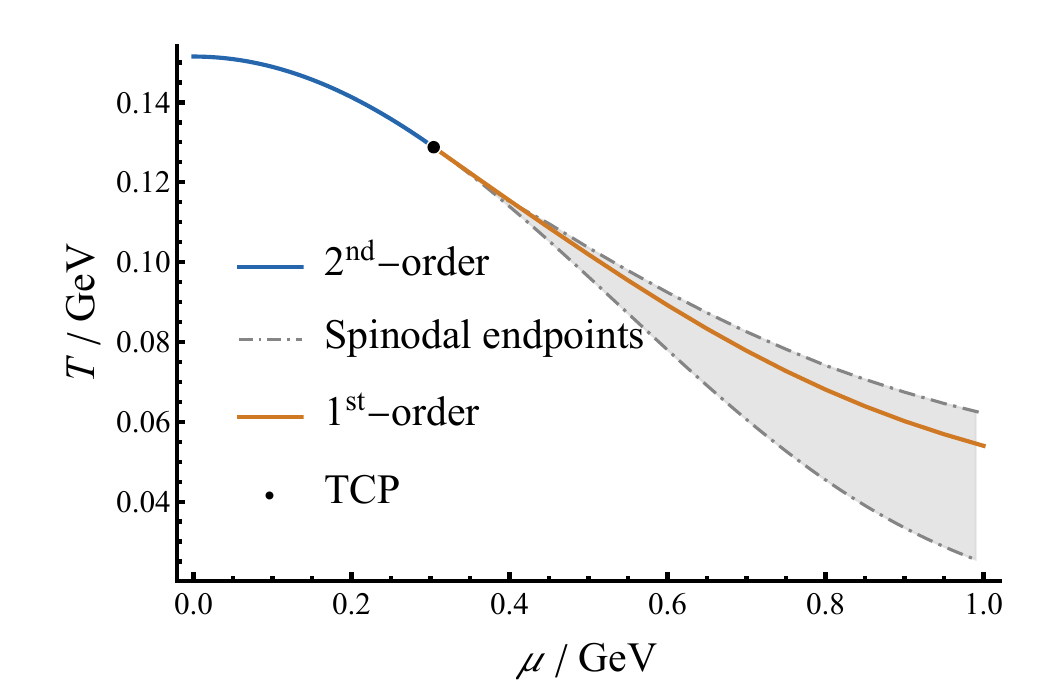}
\caption{Equilibrium chiral phase diagram in the \(T-\mu\) plane at \(m_q=0\). The second-order transition line (blue solid) terminates at the tricritical point (TCP, black dot) and converts into a first-order coexistence line (solid orange). The spinodal lines (gray dot-dashed) bound the shaded multivalued region, inside which metastable and unstable branches coexist. Temperatures and chemical potentials are in GeV.}
\label{fig:phase_diagrams}
\end{figure}

By mapping out the stable branches, coexistence lines, and spinodal endpoints, we construct the equilibrium phase diagram in the chiral limit, shown in Fig.~\ref{fig:phase_diagrams}. The blue solid line denotes the second-order transition, which terminates at a tricritical point (TCP) located at \((\mu_{\rm TCP}, T_{\rm TCP}) = (0.304, 0.128795430)\,\mathrm{GeV}\). For \(\mu > \mu_{\rm TCP}\), the transition becomes first-order, indicated by the solid orange coexistence curve. The gray dot-dashed curves denote the spinodal lines, bounding the multivalued region (shaded in gray). 

We emphasize that this clean structure, particularly the TCP, is a direct consequence of taking the exact chiral limit (\(m_q=0\)). At finite quark mass, the explicit breaking of chiral symmetry would replace the second-order line with a smooth crossover and the TCP with a critical end point (CEP). The dynamical analysis in the following sections strictly concerns this chiral limit structure. The presence of the TCP and the clean separation between the critical and spinodal points are what allow us to systematically classify the non-equilibrium dynamics.

\section{Real-Time Dynamics near the Second-Order Transition}\label{sec:second_dynamics}

Having established the equilibrium phase diagram in Sec.~\ref{sec:model}, we now shift our focus from static thermodynamics to non-equilibrium dynamics. In realistic heavy-ion collisions, the medium undergoes rapid expansion and cooling, which effectively drives the chiral condensate out of equilibrium. The relaxation of the condensate towards its equilibrium state—especially near the critical endpoints where the effective potential becomes exceedingly flat—is a genuine time-dependent process that cannot be captured by static snapshots alone. While a complete simulation of a heavy-ion collision would require coupling the condensate to an expanding hydrodynamic background, we instead focus on a homogeneous quench process at strictly fixed temperature and chemical potential. This idealized setup allows us to qualitatively isolate the intrinsic relaxation dynamics and systematically extract the fundamental dynamical laws governing the condensate evolution.

\subsection{Real-time evolution and characterization of slow roll}\label{sec:slow_roll}

To investigate this real-time evolution, we formulate an initial-value problem for the bulk scalar field. We introduce the ingoing Eddington--Finkelstein coordinate \(v\) through \(dv=dt-dz/f(z)\)~\cite{Son2002Minkowski,Chesler2014Numerical}, which is naturally adapted to the causal infalling condition at the horizon. The metric takes the form
\begin{equation}
\label{eq:ef}
ds^2=\frac{1}{z^2}\left[-f(z)\,dv^2-2\,dv\,dz+d\bm{x}^2\right].
\end{equation}
In these coordinates, the causal structure is manifest, allowing us to numerically evolve the system forward in time without encountering spurious reflections at the horizon. The time-dependent equation of motion for $\chi$ reads
\begin{equation}
\label{eq:realtime}
\begin{split}
0={}&2\partial_v\partial_z\chi
-\left(\frac{3}{z}+\Phi'\right)\partial_v\chi
-f\partial_z^2\chi\\
&+\left(\frac{3f}{z}+f\Phi'-f'\right)\partial_z\chi
+\frac{1}{z^2}\partial_\chi U.
\end{split}
\end{equation}
To study the relaxation dynamics, we initialize the system with an out-of-equilibrium radial profile:
\begin{equation}
\chi(z,0) = \sigma_i \frac{z^3}{\zeta},
\end{equation}
where \(\sigma_i\) is the initial boundary condensate. While this profile satisfies the required UV source condition, it does not satisfy the static bulk equation of motion. This mismatch introduces a genuine non-adiabatic perturbation, which excites high-frequency bulk modes and leads to an initial rapid transient stage before the system settles into a slower evolution. We have verified that the qualitative dynamical behavior is insensitive to the specific initial protocol. In particular, initializing the system with an equilibrium configuration at a different temperature (a sudden thermal quench) yields the same qualitative relaxation features \cite{Cao:2023}. Therefore, to isolate the intrinsic dynamics while maintaining computational simplicity, we adopt this boundary-profile initialization throughout this work.

We choose the integration constant in the coordinate transformation so that the ingoing time \(v\) coincides with the boundary time \(t\) at the AdS boundary (\(z=0\)). Hereafter, we simply denote the ingoing time coordinate by \(t\). The time-dependent condensate is extracted from the normalizable boundary coefficient:
\begin{equation}
\label{eq:condensate_extraction}
\sigma(t)=\zeta\lim_{z\to0}\frac{\chi(z,t)}{z^3}.
\end{equation}
This initial-value problem describes the relaxation of the condensate at fixed temperature and chemical potential, following the real-time soft-wall approach of Ref.~\cite{Cao:2023}.

\begin{figure}[h]
\centering
\includegraphics[width=0.98\columnwidth]{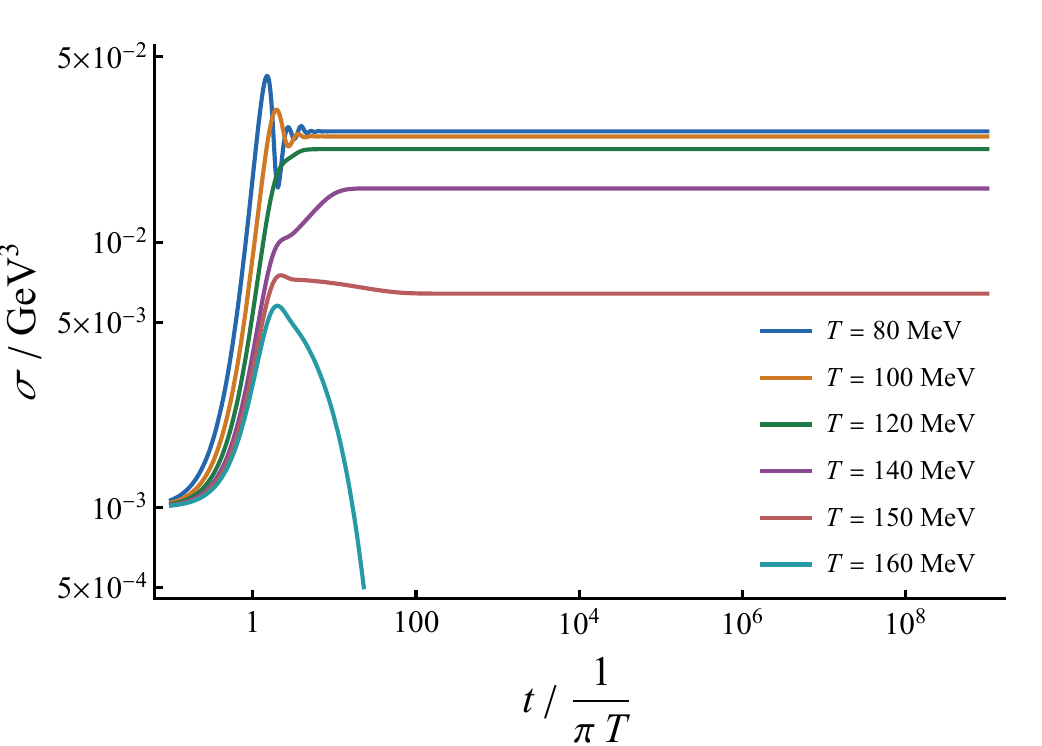}
\caption{Condensate evolution at \(\mu=0\) with initial condensate \(\sigma_i=10^{-3}\,\mathrm{GeV}^3\). The curves correspond to temperatures \(T = 80,\,100,\,120,\,140,\,150\), and \(160\,\mathrm{MeV}\), as indicated in the legend. The trajectory at \(T=160\,\mathrm{MeV}\) initially increases and then decays toward zero.}
\label{fig:sigma-t_examples}
\end{figure}

Figure~\ref{fig:sigma-t_examples} compares the real-time condensate evolution at several temperatures, starting from the same initial state with \(\sigma_i = 10^{-3}\,\mathrm{GeV}^3\) at \(\mu=0\). The selected temperatures are \(T=80,\,100,\,120,\,140,\,150\), and \(160\,\mathrm{MeV}\). For the lower temperatures \(T \le 120\,\mathrm{MeV}\) deep in the broken phase, the initial growth strongly overshoots the stationary value, followed by pronounced damped oscillations. Closer to the critical point at \(T=140\) and \(150\,\mathrm{MeV}\), the relaxation becomes much more gradual, with the \(150\,\mathrm{MeV}\) trajectory exhibiting an extended smooth approach after a slight initial overshoot. Above the critical temperature $T_c(\mu=0)$, the trajectory at \(T=160\,\mathrm{MeV}\) initially increases to a transient peak and subsequently decays toward the symmetric state (zero condensate). This decay ultimately falls below the displayed range of the plot.

\begin{figure}[tbp]
\centering
% (a)
\begin{overpic}[width=0.94\columnwidth]
{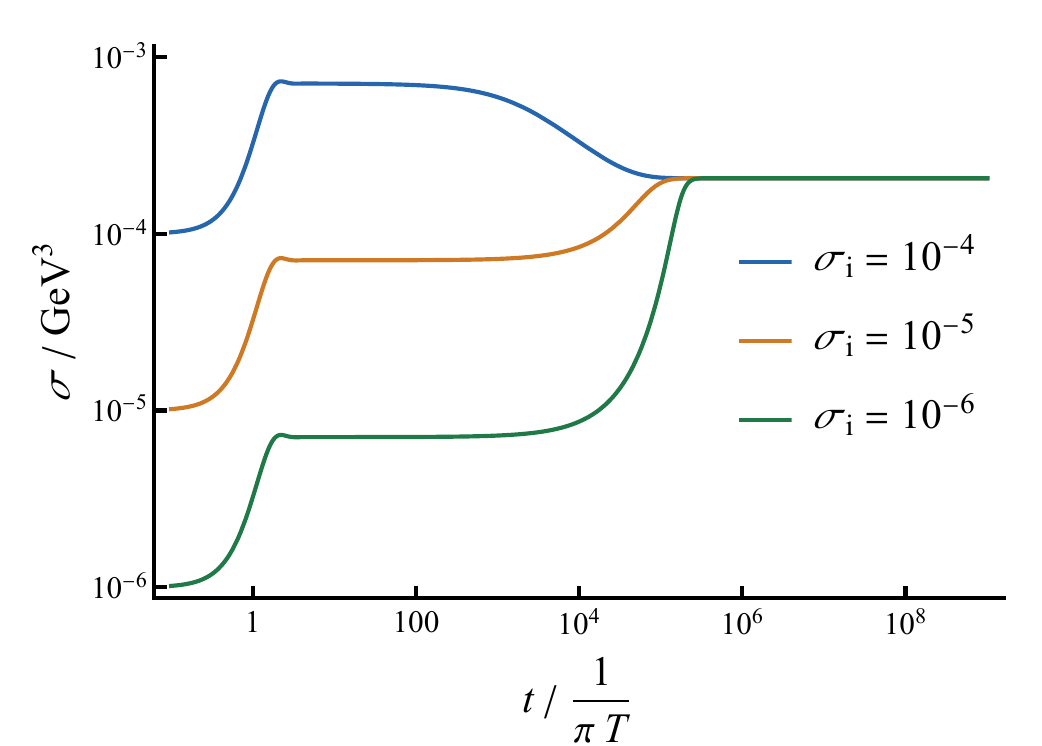}
\put(50,65){\color{black}\small (a) $T<T_c$}
\end{overpic}
% (b)
\begin{overpic}[width=0.94\columnwidth]
{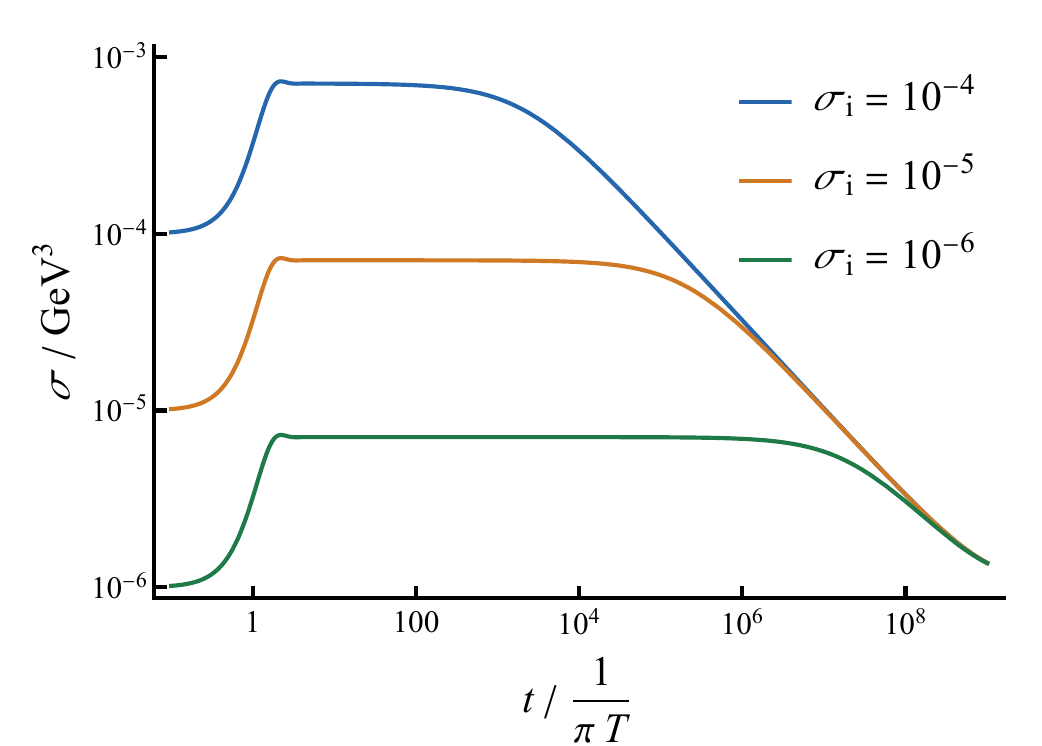}
\put(50,65){\color{black}\small (b) $T=T_c$}
\end{overpic}
% (c)
\begin{overpic}[width=0.94\columnwidth]
{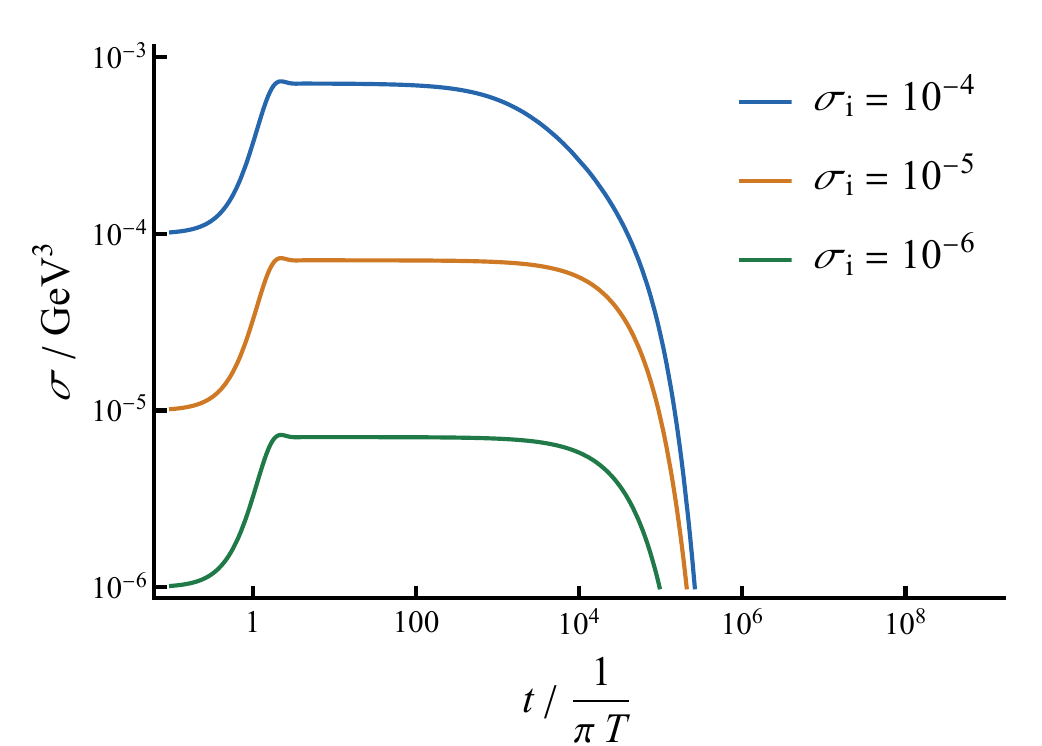}
\put(50,65){\color{black}\small (c) $T>T_c$}
\end{overpic}
\caption{Condensate evolution near the second-order transition at \(\mu=0\) for different initial condensates. From top to bottom, the three panels correspond to the relative temperature deviations \(\epsilon = (T-T_c)/T_c = -10^{-5}, 0\), and \(+10^{-5}\), respectively. (a) just below \(T_c\) (\(\epsilon < 0\)), (b) exactly at \(T_c\) (\(\epsilon = 0\)), and (c) just above \(T_c\) (\(\epsilon > 0\)). In each panel, the blue, orange, and green curves denote the initial condensates \(\sigma_i = 10^{-4}\), \(10^{-5}\), and \(10^{-6}\,\mathrm{GeV}^3\), respectively. Following an initial transient, the trajectories develop an intermediate stage of slow evolution. All three trajectories in panel (c) ultimately approach zero.}
\label{fig:sigma_tc_initial}
\end{figure}

An interesting feature emerges as the temperature approaches \(T_c\). Focusing on the zero chemical potential case (\(\mu=0\)), we quantify the temperature deviation via \(\epsilon = (T-T_c)/T_c\). Figure~\ref{fig:sigma_tc_initial} presents the evolution for \(\epsilon = -10^{-5}, 0\), and \(+10^{-5}\) (i.e., \(T \simeq 0.151514130, 0.151515645\), and \(0.151517160\,\mathrm{GeV}\)), each initialized with three different condensates \(\sigma_i = 10^{-4}, 10^{-5}\), and \(10^{-6}\,\mathrm{GeV}^3\). 

Despite the different final fates-relaxing to a nonzero equilibrium below \(T_c\), approaching a universal decay curve at \(T_c\), and decaying toward zero above \(T_c\)---all trajectories share a common feature: they develop a pronounced intermediate stage between the initial transient and late-time relaxation, during which the condensate evolves extremely slowly. This qualitative behavior persists across the initial conditions and temperatures considered, while the condensate amplitude and the duration of the intermediate stage vary.

The emergence of such a long-lived intermediate stage has been widely discussed in quantum field theory in the context of prethermalization~\cite{Berges2004Prethermalization}, and similar quasi-steady plateaus have been identified in holographic chiral dynamics~\cite{Cao:2023,Zheng:2025,Giannuzzi:2025,Tang:2025}. However, the internal dynamical structure of this regime remains unresolved: is it a strictly static plateau, or a continuous dynamical drift? To answer this, we need to zoom in on the evolution within this interval.

\begin{figure}[bp]
\centering
{\small (a)\par}
\includegraphics[width=0.96\columnwidth]{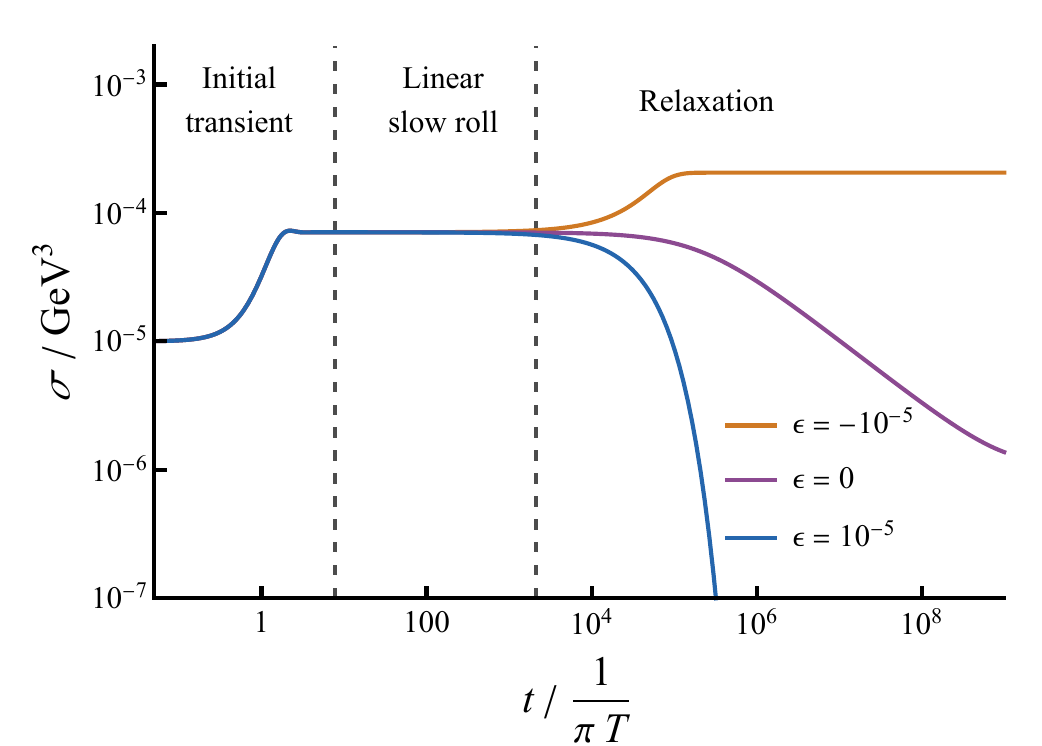}
\par\smallskip
{\small (b)\par}
\includegraphics[width=0.96\columnwidth]{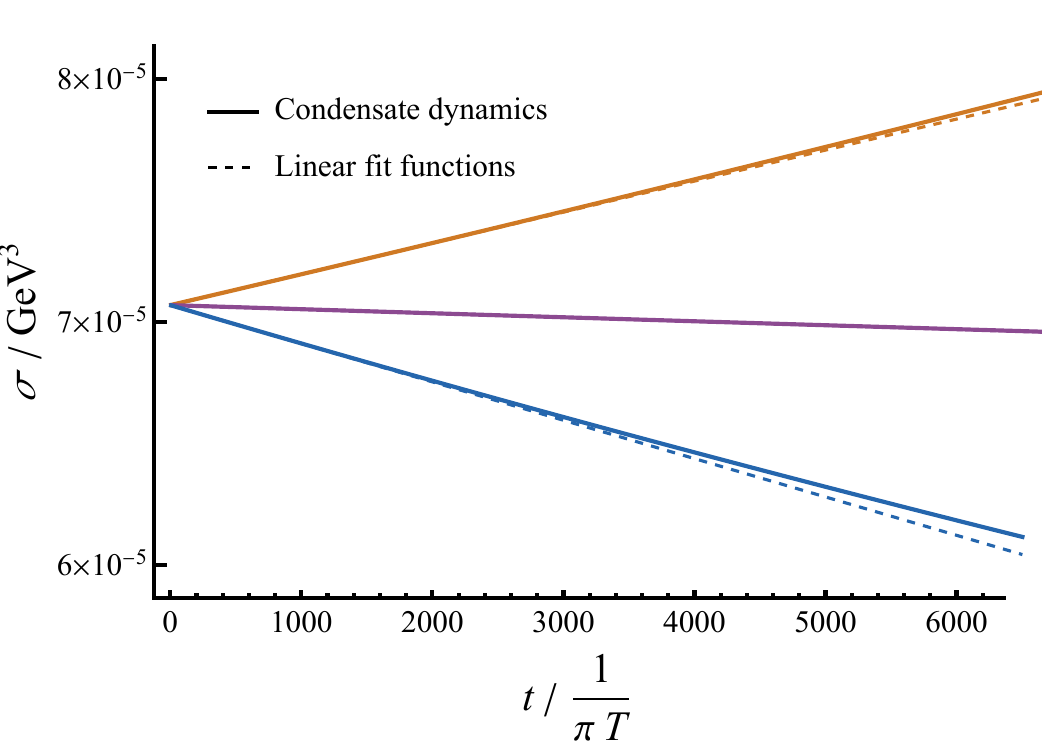}
\caption{Slow evolution of the chiral condensate near $T_c$ at $\mu=0$ with initial condensate \(\sigma_i=10^{-5}\,\mathrm{GeV}^3\), for $\epsilon=(T-T_c)/T_c=-10^{-5},0,10^{-5}$. (a) The initial transient, intermediate slow-roll stage, and subsequent relaxation, with vertical dashed lines marking their approximate separation. (b) Approximately linear evolution within the intermediate regime. Solid curves show the numerical trajectories, and dotted lines show straight-line approximations.}
\label{fig:early_slow_evolution}
\end{figure}

To address this question, we examine the detailed trajectory within the intermediate interval. Figure~\ref{fig:early_slow_evolution}(a) shows the full trajectories for the same three temperature detunings. We can clearly identify three distinct stages: an initial rapid transient, an intermediate slow-roll stage, and a subsequent relaxation toward equilibrium, with their approximate boundaries marked by vertical dashed lines. 

While the intermediate stage appears as a quasi-plateau on a logarithmic scale, a closer inspection on a linear scale reveals its internal structure. As shown in Fig.~\ref{fig:early_slow_evolution}(b), during this intermediate stage, the condensate undergoes a continuous slow roll with an approximately constant drift rate over an extended time interval. We refer to this novel dynamical regime as the ``linear slow roll''. This behavior is quantitatively captured by the linear approximation:
\begin{equation}
\label{eq:linear_approximation}
\sigma(t)\simeq\sigma_{\mathrm{linear}}(t)
=\sigma(t_L)+K(t-t_L),\quad t\in[t_L,t_R],
\end{equation}
where \(t_L\) and \(t_R\) denote the beginning and end of the identified linear interval, respectively, and \(K\) is the constant drift rate. We refer to the condensate value at the interval entrance, \(\sigma_L \equiv \sigma(t_L)\), as the entrance condensate. The extraction of the interval boundaries relies on a strict linearity criterion, the details of which are deferred to Appendix~\ref{app:duration_derivation}.

Because the drift rate \(K\) is extremely small, this slow evolution macroscopically manifests as a quasi-steady state, consistent with the prethermalization plateau identified in earlier studies. However, our analysis reveals the fine dynamical structure of this regime: rather than being strictly static, the system undergoes a continuous linear drift.

For the initial conditions considered, the duration of this slow-roll stage \(\Delta t = t_R - t_L\) becomes parametrically extended as the temperature approaches \(T_c\) (\(\epsilon \to 0\)). Remarkably, as the entrance condensate decreases, this linear regime can persist for a parametrically long time, reaching up to \(10^3\) in units of \(1/(\pi T)\) (e.g., for the three temperatures shown in Fig.~\ref{fig:early_slow_evolution}, the extracted intervals \([t_L, t_R]\) are approximately \([6.8, 1.9\times10^3]/(\pi T)\), \([9.5, 3.5\times10^3]/(\pi T)\), and \([6.8, 9.1\times10^2]/(\pi T)\) for \(\epsilon = -10^{-5}, 0\), and \(+10^{-5}\), respectively). Furthermore, the sign of the drift slope \(K\) directly reflects the underlying potential gradient: it is positive for \(\epsilon < 0\) (driving condensate growth), weakly negative at \(\epsilon = 0\), and more negative for \(\epsilon > 0\) (driving decay). This clear dependence on the temperature deviation motivates us to seek a reduced dynamical equation that can analytically explain the origin of this nearly constant evolution rate and its sensitivity to \(\epsilon\).

It is also worth emphasizing that, as indicated by the constant slope in Fig.~\ref{fig:early_slow_evolution}(b), this intermediate linear drift is a distinct dynamical regime. Unlike late-time relaxation, where the velocity continuously decays toward equilibrium, or critical slowing down, where the relaxation time diverges at a single point, the system here undergoes a continuous, nearly constant-velocity drift over a broad time window. 

To uncover the microscopic origin of this constant drift and to analytically justify its longevity, we must look beyond the boundary condensate and examine the full 5D bulk dynamics. It is within this higher-dimensional geometry that the underlying mechanism for the ultra-long slow roll is hidden.

\subsection{The slow configuration space and effective Ginzburg-Landau equation}\label{sec:reduce}

To uncover the hidden mechanism within the 5D bulk geometry, we must understand how non-equilibrium dynamics operates in a strongly coupled system. In the holographic framework, the bulk field contains an infinite tower of modes. However, the causal event horizon acts as an ideal absorber: any high-frequency (non-adiabatic) excitation falling into the black hole is completely dissipated. Consequently, after the initial transient (\(t \gtrsim 2/(\pi T)\)), all fast modes are exponentially damped. The subsequent evolution is entirely governed by the lowest-lying (zero-frequency at $T_c$) mode, which is dictated by the evolving boundary condensate \(\sigma(t)\).

In the language of dynamical systems, the bulk field is adiabatically slaved to a one-dimensional slow configuration space parameterized solely by the instantaneous boundary condensate \(\sigma(t)\). Because the boundary condensate itself evolves on a time scale much longer than the thermal relaxation time \(1/(\pi T)\), the bulk profile at any instant is effectively a frozen snapshot of this space. This geometric slaving is what guarantees that a single scalar function \(\sigma(t)\) can capture the full complexity of the 5D real-time evolution.

This dynamical slaving admits a transparent macroscopic analogy. Consider a gas in a piston subjected to an extremely strong friction (the overdamped limit). After a violent initial transient, the friction forces the piston to move so slowly that the gas pressure is instantaneously determined by the volume, \(P=P(V)\). Although the overall expansion is strongly dissipative and irreversible, the instantaneous state is slaved to a single external parameter.

In our holographic setup, the bulk QNM excitations play the role of the violent initial transient, while the causal event horizon acts as the strong friction. Crucially, the 5D bulk field does not reach local thermodynamic equilibrium; its
radial profile is intrinsically inhomogeneous. Instead, after a fast relaxation time \(t \sim 2/(\pi T)\), the high-frequency modes are completely absorbed, and the entire bulk profile is dynamically slaved to an instantaneous steady state uniquely determined by the boundary condensate.

\begin{figure}[tbp]
\centering
\includegraphics[width=0.98\columnwidth]{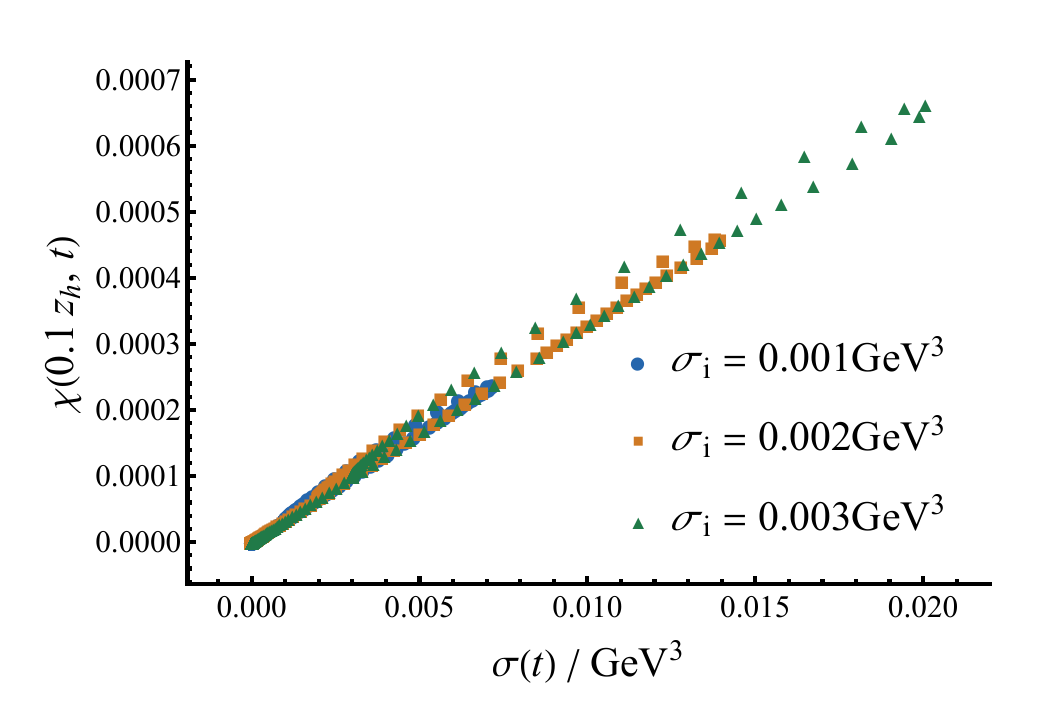}
\includegraphics[width=0.98\columnwidth]{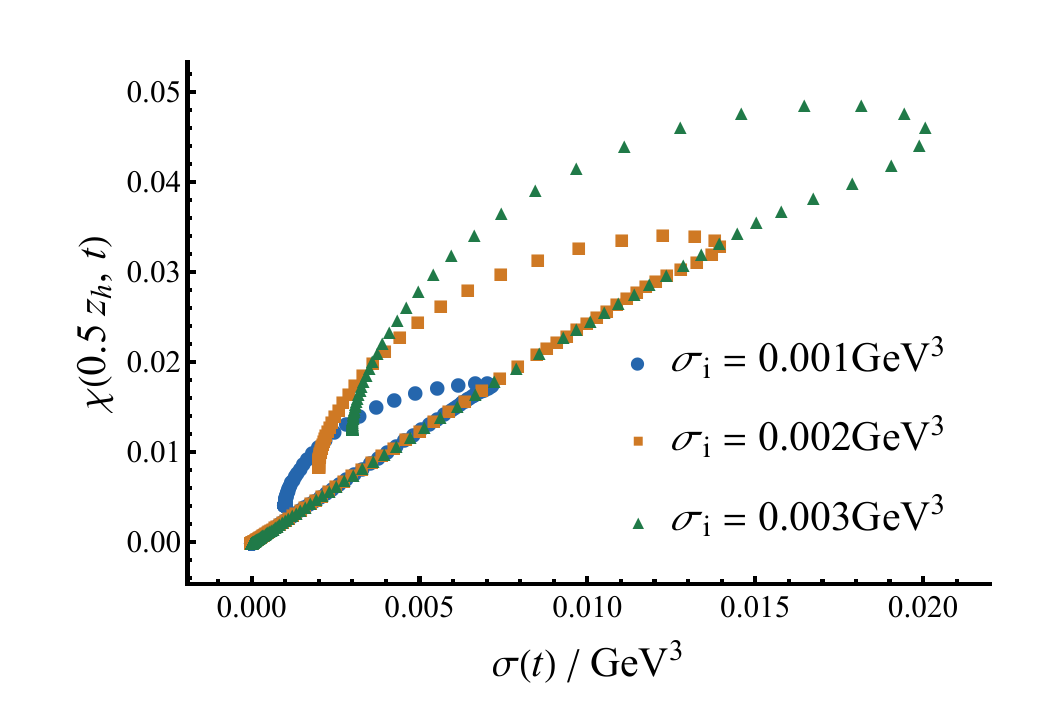}
\includegraphics[width=0.98\columnwidth]{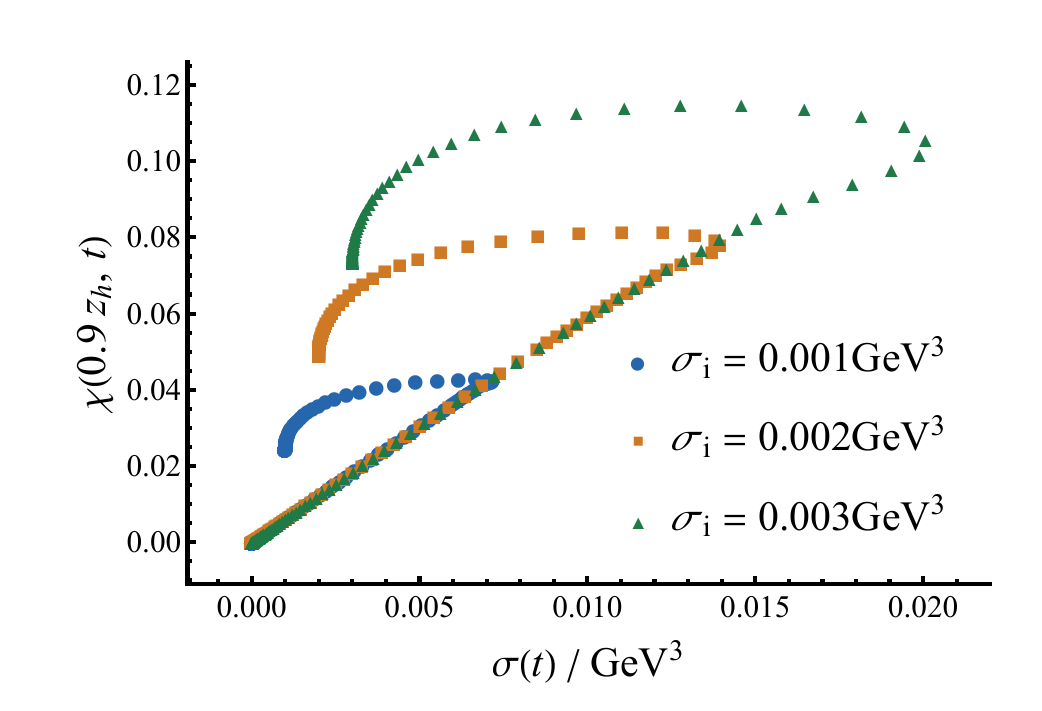}
\caption{Bulk-field trajectories in the \((\sigma(t), \chi(z,t))\) plane at \(T=T_c\) and \(\mu=0\). The three panels from top to bottom correspond to radial slices at \(z=0.1z_h\), \(0.5z_h\), and \(0.9z_h\), respectively. In each panel, trajectories originating from different initial condensates \(\sigma_i = (0.001, 0.002, 0.003)\,\mathrm{GeV}^3\) rapidly collapse onto a single universal branch after the transient at \(t \simeq 2/(\pi T)\), demonstrating the dynamical slaving to the slow configuration space.}
\label{fig:chi_sigma_critical}
\end{figure}

We now verify this physical picture with our full numerical evolution. Figure~\ref{fig:chi_sigma_critical} displays the bulk field \(\chi(z,t)\) at three representative radial positions (\(z=0.1z_h, 0.5z_h\), and \(0.9z_h\)) as a function of the instantaneous condensate \(\sigma(t)\). Despite starting from different initial transients, the trajectories at each radial position rapidly converge and collapse onto a single common branch after \(t \simeq 2/(\pi T)\). This remarkable waveform collapse provides a direct geometric confirmation: the high-frequency bulk modes are indeed dissipated, and the entire 5D field is dynamically slaved to the slow configuration space.

\begin{figure}[htbp]
\centering
\includegraphics[width=0.98\columnwidth]{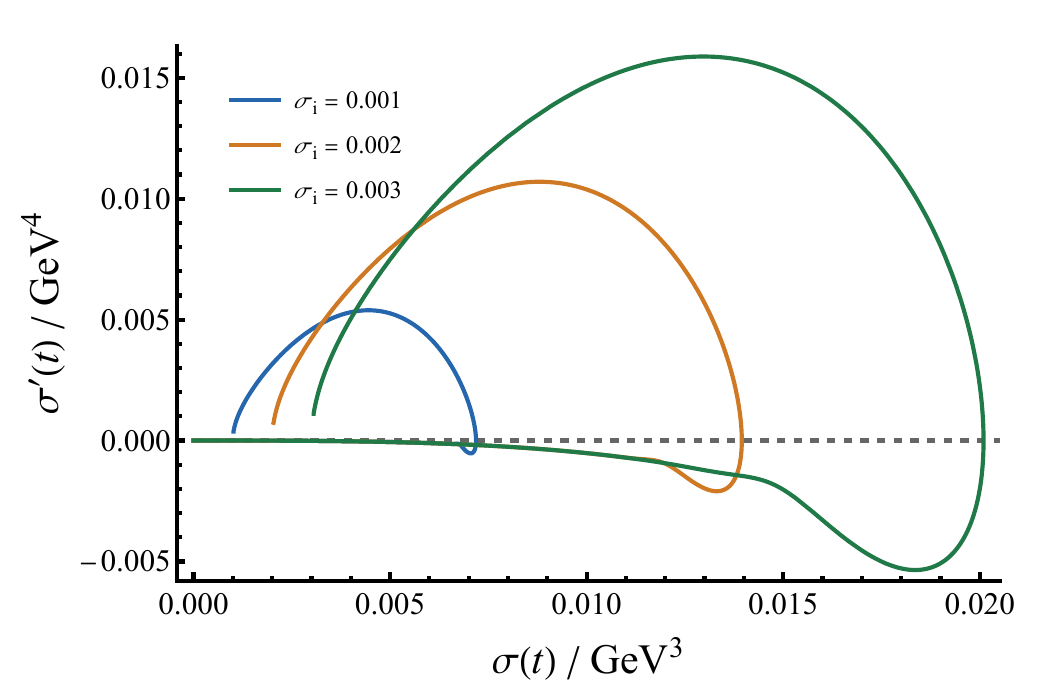}
\caption{Condensate drift as a function of the condensate at \(T=T_c\) and
\(\mu=0\) for \(\sigma_i=(0.001,\,0.002,\,0.003)\,\mathrm{GeV}^3\). Colored
curves show the full numerical trajectories in the \((\sigma,\dot\sigma)\) plane.
The early transients exhibit initial-condition-dependent loops. After the
transient, all trajectories collapse onto a universal attractor curve.}
\label{fig:condensate_drift_attractor}
\end{figure}

To further verify this dynamical slaving, we examine the phase-space evolution. Figure~\ref{fig:condensate_drift_attractor} shows the trajectories of the condensate drift \(\dot{\sigma}(t)\) as a function of the instantaneous condensate \(\sigma(t)\) for three different initial values
\(\sigma_i = (0.001,\, 0.002,\, 0.003)\,\mathrm{GeV}^3\) at \(T=T_c\) and \(\mu=0\). The early-time evolution exhibits large transient loops that depend strongly on the initial conditions, driven by non-adiabatic bulk excitations. However, after the initial transient, all trajectories rapidly collapse onto a single universal curve in the \((\sigma, \dot{\sigma})\) plane. The system completely loses memory of its initial state, and its drift velocity is uniquely dictated by the instantaneous condensate value.

This geometric collapse in the phase space provides a direct manifestation of the holographic attractor mechanism~\cite{Heller:2011ju,Heller:2013fn}. Much like the hydrodynamic attractors where the late-time stress tensor is determined by a few fluid variables, the order parameter here is dynamically confined to a one-dimensional invariant manifold parameterized solely by \(\sigma(t)\). Therefore, not only the bulk field profile but also its time derivative is dynamically slaved to the evolving boundary condensate.

This observation motivates us to assume that after the transient, the bulk field is adiabatically slaved to the slow configuration space. Therefore, the field profile can be parameterized entirely by the instantaneous boundary condensate:
\begin{equation}\label{eq:invariant_manifold_ansatz}
\chi(z,\sigma,t) \simeq \bar{\chi}\bigl(z,\sigma(t)\bigr).
\end{equation}
This form of adiabatic expansion was previously adopted in holographic quenches
across critical points~\cite{Basu:2011ft}. Here, however, the key feature is that
the holographic attractor mechanism ensures this slaving remains robust over an
exceptionally long intermediate-time window.

Substituting this adiabatic ansatz into the real-time equation of motion \eqref{eq:realtime}, the time derivative is evaluated via the chain rule as \(\partial_t\chi = \partial_\sigma\bar{\chi} \cdot \dot{\sigma}\). This yields:
\begin{equation}
\dot{\sigma}\,\mathcal{D}\!\left[\partial_\sigma\bar{\chi}\right]
=R[\bar{\chi}],
\end{equation}
where the derivative operator \(\mathcal{D}\) and the spatial operator \(R[\bar{\chi}]\) are defined as
\begin{align}
\mathcal{D}&=2\partial_z-\left(\frac{3}{z}+\Phi'\right), \label{eq:time_derivative_operator}\\
R[\bar{\chi}]&=f\,\partial_z^2\bar{\chi}
-\left(\frac{3f}{z}+f\Phi'-f'\right)\partial_z\bar{\chi}
-\frac{1}{z^2}\partial_\chi U. \label{eq:radial_evolution_operator}
\end{align}

Since the right-hand side of the above equation depends only on the spatial profile \(\bar{\chi}\) and the parameter \(\sigma\), the drift velocity \(\dot{\sigma}\) must be a function of \(\sigma\) alone. We therefore define the reduced drift \(\mathcal{F}(\sigma) \equiv \dot{\sigma}\). The dynamics of the condensate is thus governed by the invariance equation:
\begin{equation}\label{eq:invariance_equation}
\mathcal{F}(\sigma)\,\mathcal{D}\!\left[\partial_\sigma\bar{\chi}\right]
=R[\bar{\chi}].
\end{equation}
To uniquely solve this equation, we must impose the appropriate boundary conditions. Near the UV boundary (\(z \to 0\)), the fixed source condition determines the asymptotic behavior \(\bar{\chi}(z) \sim \sigma z^3 / \zeta\). At the IR event horizon (\(z=z_h\)), the physical field must be perfectly regular. However, since \(f(z_h)=0\), the highest derivative term \(f(z)\partial_z^2\bar{\chi}\) vanishes, causing the differential equation to degenerate. To prevent a divergence in \(\partial_z^2\bar{\chi}\), the entire equation at \(z=z_h\) must reduce to an exact algebraic constraint:
\begin{equation}
\mathcal{F}(\sigma)\,\mathcal{D}\!\left[\partial_\sigma\bar{\chi}\right]_{z=z_h}
= f'(z_h)\partial_z\bar{\chi}\Big|_{z=z_h} - \frac{1}{z_h^2}\partial_\chi U(\bar{\chi}_{z_h}).
\end{equation}
This horizon regularity condition is mathematically equivalent to the causal infalling condition in the real-time evolution. Combined with the UV boundary condition, it provides a closed boundary-value problem that uniquely fixes the radial profile \(\bar{\chi}(z, \sigma)\) and determines the reduced drift \(\mathcal{F}(\sigma)\) self-consistently.

Crucially, Eq.~\eqref{eq:invariance_equation} serves as the exact microscopic realization of the Ginzburg-Landau (GL) dynamics in the holographic bulk. In the conventional phenomenological GL theory, the order parameter evolution is governed by the overdamped equation \(\dot{\sigma} = -\Gamma \partial_\sigma V_{\text{eff}}(\sigma)\), where both the effective potential \(V_{\text{eff}}\) and the kinetic coefficient \(\Gamma\) are introduced as phenomenological inputs. In our holographic approach, however, the GL equation naturally emerges as the boundary projection of the invariance equation. By matching the asymptotic expansion near \(z \to 0\), the bulk equation projects onto the boundary as a purely algebraic relation, uniquely extracting the reduced drift \(\mathcal{F}(\sigma) = -\Gamma \partial_\sigma V_{\text{eff}}(\sigma)\). Here, the kinetic coefficient \(\Gamma\) arises from the horizon dissipation, and the effective potential \(V_{\text{eff}}\) is self-consistently generated by the bulk geometry. Thus, the holographic bulk provides the fundamental microscopic origin for the phenomenological GL parameters.

At $\mu=0$, the 5D bulk scalar potential reduces to \(U(\chi)=-\frac{3}{2}\chi^2+v_{40}\chi^4+v_{60}\chi^6\). The bulk dynamics are invariant under \(\chi\to-\chi\). We choose a configuration parameterization that respects this symmetry and expand the field profile and the reduced drift in odd powers of the condensate:
\begin{align}
\bar{\chi}(z,\sigma) &= \sigma\chi_1(z) + \sigma^3\chi_3(z) + O(\sigma^5), \label{eq:manifold_amplitude_expansion}\\
\mathcal{F}(\sigma) &= c_1\sigma + c_3\sigma^3 + O(\sigma^5). \label{eq:reduced_drift_expansion}
\end{align}

The radial functions are normalized so that $\sigma$ remains the physical boundary condensate. Substituting these expansions into Eq.~\eqref{eq:invariance_equation} determines the drift coefficients through radial boundary-value problems, as detailed in Appendix~\ref{app:finite_temperature_reduction}. The linear coefficient $c_1$ is the eigenvalue of the mode connected to the critical zero mode and vanishes at $T_c$.

\subsection{Reduced dynamics and numerical verification near \texorpdfstring{\(T_c\)}{Tc}}\label{sec:cubic_verification}

At $T_c$, the linear coefficient vanishes, and the leading condensate evolution is governed by the cubic term. The calculation gives $c_3(T_c)\simeq-476.58\,\mathrm{GeV}^{-5}$. Integrating the cubic evolution equation with $\sigma(t_0)=\sigma_0$ yields
\begin{equation}
\label{eq:critical_condensate_solution}
\sigma(t)=
\frac{\sigma_0}
{\sqrt{1-2c_3(T_c)\sigma_0^2(t-t_0)}}.
\end{equation}

\begin{figure}
\centering
\includegraphics[width=0.98\columnwidth]{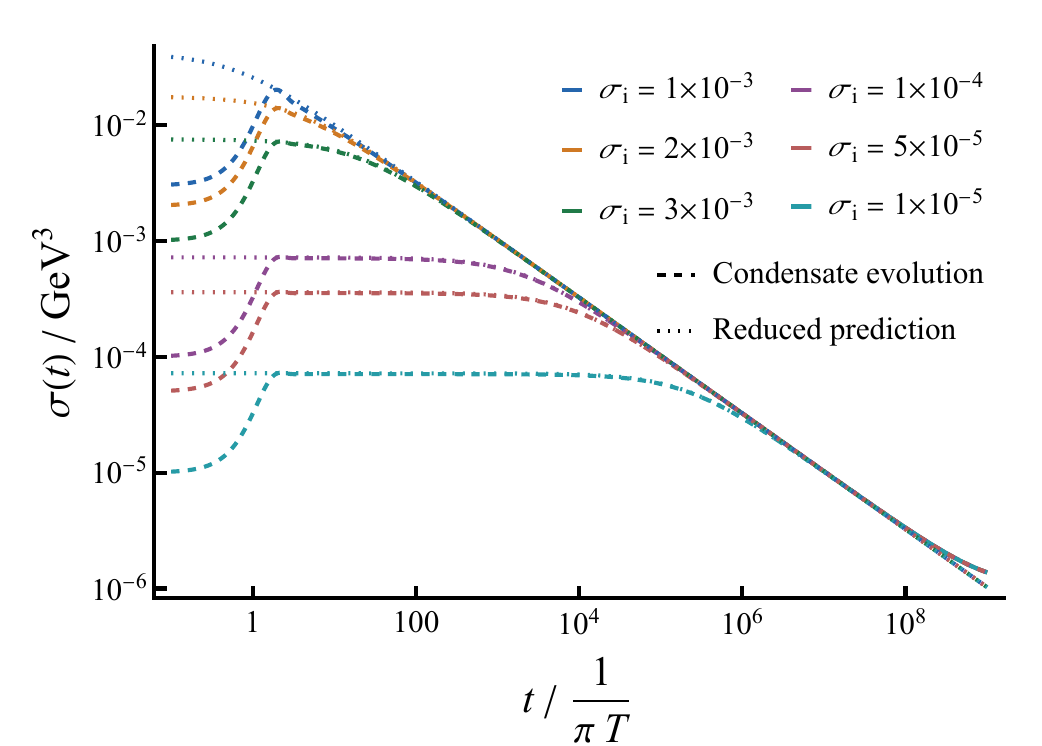}
\caption{Condensate evolution (dashed) and analytical predictions from Eq.~\eqref{eq:critical_condensate_solution} (dotted) at \(T_c\) and \(\mu=0\), with six \(\sigma_i\). The predictions use \(\sigma_0=\sigma(t_0)\) from the full numerical solution and \(t_0 \simeq 2/(\pi T)\). The coefficients \(c_1(T)\) and \(c_3(T)\) are computed analytically from the bulk geometry, as listed in Table~\ref{tab:second_order_reduced_coefficients}.}
\label{fig:critical_reduced_prediction}
\end{figure}

Figure~\ref{fig:critical_reduced_prediction} compares this prediction with the full numerical evolution for six initial condensates. Since the bulk trajectories approach a common slow configuration by $t\simeq2/(\pi T)$, as shown in Fig.~\ref{fig:chi_sigma_critical}, we initialize each prediction at $t_0\simeq2/(\pi T)$ using $\sigma_0=\sigma(t_0)$ from the corresponding numerical trajectory. The same independently calculated coefficient $c_3(T_c)$ is used for all six trajectories. The predictions reproduce the intermediate slow evolution and its subsequent crossover to decay.

As shown in Fig.~\ref{fig:critical_reduced_prediction}, the analytical predictions closely follow the numerical trajectories over a broad time interval, capturing the dependence of the intermediate evolution on the initial condensate and the subsequent crossover to decay. At the latest displayed times, the numerical condensates lie slightly above the cubic predictions. This deviation is consistent with a small residual positive linear drift caused by a slight offset of the numerical temperature from the exact critical point, whereas the analytical prediction imposes $c_1=0$.

We next extend the calculation to temperatures slightly away from \(T_c\). The coefficients $c_1(T)$ and $c_3(T)$ are determined from the constrained radial problems in Appendix~\ref{app:finite_temperature_reduction}, following the mode continuously connected to the critical zero mode. Their values are summarized in Table~\ref{tab:second_order_reduced_coefficients}. Two important features emerge from the table: First, the linear coefficient \(c_1(T)\) scales approximately linearly with the temperature deviation \(\epsilon\), satisfying \(c_1(T) \simeq -2.02 \epsilon \mathrm{GeV}\). Second, the cubic coefficient \(c_3(T)\) remains nearly constant near \(c_3 = -476.58\mathrm{GeV}^{-5}\) across the entire sampled range. This indicates that the linear restoring force is directly controlled by the temperature deviation from criticality, while the leading non-linear dynamics encoded in the bulk geometry is remarkably robust.

\begin{table}[htbp]
\centering
\caption{Reduced drift coefficients near the second-order critical temperature, with \(\epsilon=(T-T_c)/T_c\). The coefficient \(c_1(T)\) is in units of \(\mathrm{GeV}\), and \(c_3(T)\) is in units of \(\mathrm{GeV}^{-5}\).}
\label{tab:second_order_reduced_coefficients}
\renewcommand{\arraystretch}{1.15}
\begin{tabular*}{0.98\columnwidth}{@{\extracolsep{\fill}}ccc@{}}
\hline
\(\epsilon\) & \(c_1(T)\) & \(c_3(T)\) \\
\hline
\(-10^{-1}\) & \( 2.04\times10^{-1}\)  & \(-486.10\) \\
\(-10^{-3}\) & \( 2.02\times10^{-3}\)  & \(-476.63\) \\
\(-10^{-5}\) & \( 2.02\times10^{-5}\)  & \(-476.58\) \\
\(0 \)       & \( 0\)                 & \(-476.58\) \\
\(+10^{-5}\) & \(-2.02\times10^{-5}\) & \(-476.58\) \\
\(+10^{-3}\) & \(-2.02\times10^{-3}\) & \(-476.54\) \\
\(+10^{-1}\) & \(-2.01\times10^{-1}\) & \(-474.91\) \\
\hline
\end{tabular*}
\end{table}

Retaining terms through cubic order in Eq.~\eqref{eq:reduced_drift_expansion}, we obtain the analytical solution
\begin{equation}
\label{eq:finite_temperature_condensate_solution}
\sigma(t)=
\frac{\sigma_0 e^{c_1(t-t_0)}}
{\sqrt{1-\dfrac{c_3}{c_1}\sigma_0^2
\left[e^{2c_1(t-t_0)}-1\right]}},
\qquad c_1\neq0,
\end{equation}
where the coefficients are evaluated at the chosen temperature. Its \(c_1\to0\) limit smoothly recovers the critical solution Eq.~\eqref{eq:critical_condensate_solution}. Physically, the non-zero \(c_1\) introduces a direct competition between the linear and cubic terms: below \(T_c\) (\(c_1 > 0\)), the linear term initially drives the growth while the cubic term eventually curbs it; above \(T_c\) (\(c_1 < 0\)), both terms drive the condensate toward zero, with the linear term dominating the late-time decay.

\begin{figure}[htbp]
\centering
\includegraphics[width=0.98\columnwidth]{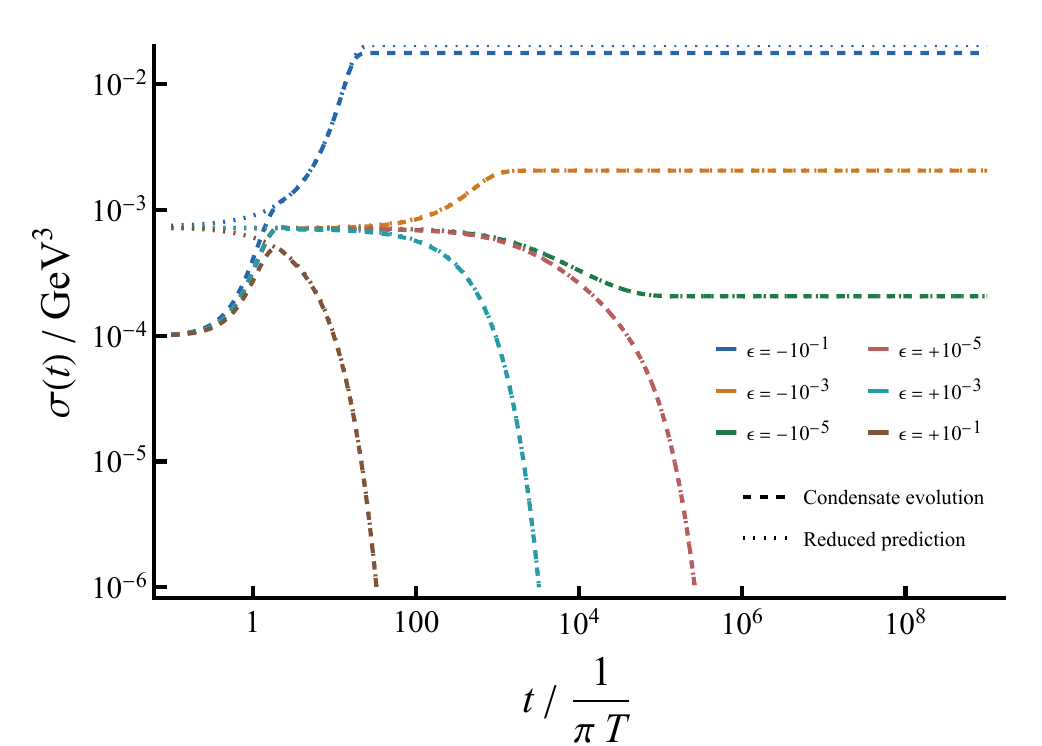}
\caption{Condensate evolution (dashed) and predictions from Eq.~\eqref{eq:finite_temperature_condensate_solution} (dotted) at temperatures near \(T_c\) and \(\mu=0\), with \(\sigma_i=10^{-4}\,\mathrm{GeV}^3\). The predictions use \(\sigma_0=\sigma(t_0)\) from the full numerical solution and \(t_0 \simeq 2/(\pi T)\), with \(c_1(T)\) and \(c_3(T)\) determined from Table~\ref{tab:second_order_reduced_coefficients}.}
\label{fig:finite_temperature_prediction}
\end{figure}

Figure~\ref{fig:finite_temperature_prediction} compares these predictions with the full numerical evolution at representative temperatures near \(T_c\). For each trajectory, we match \(\sigma_0=\sigma(t_0)\) to the full numerical solution at a reference time after the initial transient. At each temperature, the coefficients \(c_1(T)\) and \(c_3(T)\) are shared by all initial conditions. The excellent agreement between the analytic solutions and the full 5D numerical evolution further validates the slow configuration reduction and the robustness of the effective GL description.

In summary, the holographic slow configuration space reduction successfully maps the 5D real-time dynamics onto a boundary first-order GL equation. At the second-order critical point, the emergence of the zero mode directly leads to the ``double-zero'' condition, eliminating the linear restoring force and giving rise to a long-lived linear slow roll. Furthermore, our extension to \(T \neq T_c\) demonstrates that the framework remains robust even when the zero mode is broken. 

However, the second-order point is a symmetric point where the effective potential is strictly even in \(\sigma\). It is natural to ask how this dynamical mechanism generalizes to a first-order phase transition, which possesses a fundamentally different structure. In the next section, we extend our analysis to the two spinodal endpoints of the first-order transition, and we will show that the distinct local geometries at these endpoints—a symmetric zero-condensate point versus a non-zero minimum-barrier merger—will dictate fundamentally different drift laws.

\section{Linear Slow Roll Near the Spinodal Endpoints of the First-Order Phase Transition}\label{sec:first_order_dynamics}

\subsection{Slow evolution near the Spinodal Endpoints}

Having established the dynamical mechanism at the symmetric second-order critical point, we now extend our analysis to the first-order phase transition. Unlike the second-order point where the effective potential is strictly even, the first-order transition is bounded by two distinct spinodal endpoints. At \(\mu=0.6\,\mathrm{GeV}\), the equilibrium condensate exhibits a multivalued structure: the lower spinodal \(\Tm\) marks the limit of the zero-condensate metastable state, while the upper spinodal \(\Tp\) represents the merger of a non-zero minimum with the intervening barrier.

To isolate the intrinsic slow dynamics without the complication of competing metastable equilibria, we perform quenches to final temperatures just outside the spinodal interval. We select three temperature detunings \(|\epsilon| = |T - T_{\mathrm{sp}}^{\pm}|/T_{\mathrm{sp}}^{\pm} = 10^{-5}, 10^{-4}\), and \(10^{-3}\) for both endpoints. For the lower spinodal (\(T < \Tm\)), we initialize the system with small initial condensates \(\sigma_i = (10^{-6}, 10^{-5}, 10^{-4})\,\mathrm{GeV}^3\). For the upper spinodal (\(T > \Tp\)), we use larger initial condensates \(\sigma_i = (0.005, 0.006, 0.007)\,\mathrm{GeV}^3\). 

\begin{figure}[t]
\centering
{\small $(a)\ T < \Tm$\par}
\includegraphics[width=0.98\columnwidth]{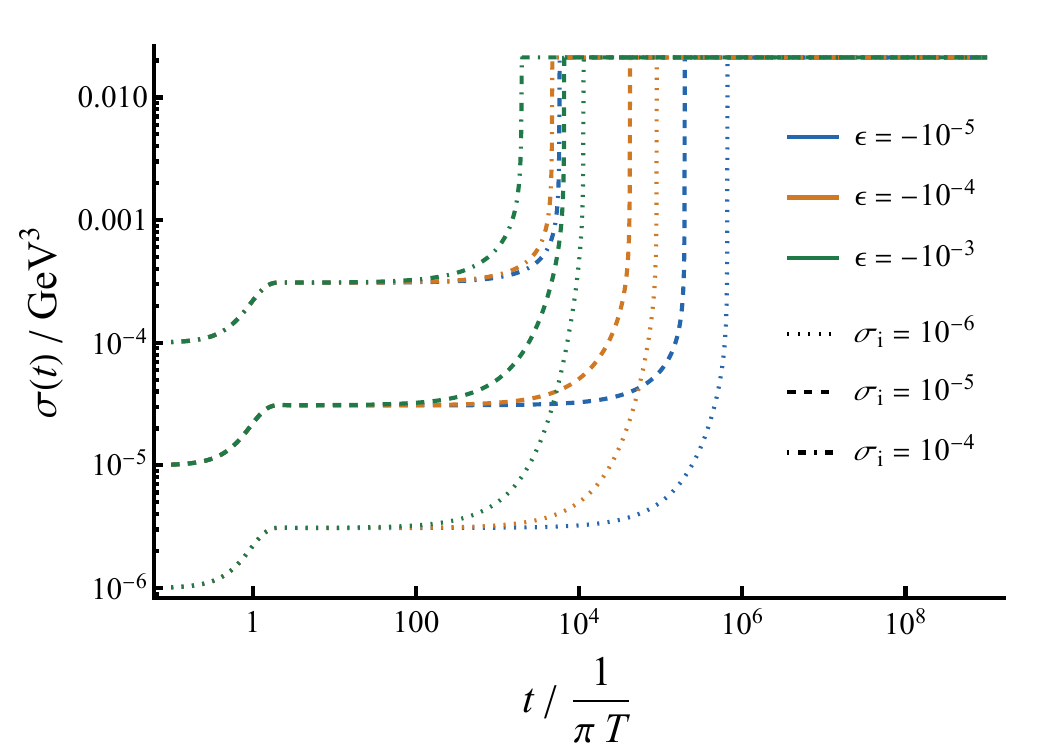}
\medskip
{\small $(b)\ T > \Tp$\par}
\includegraphics[width=0.98\columnwidth]{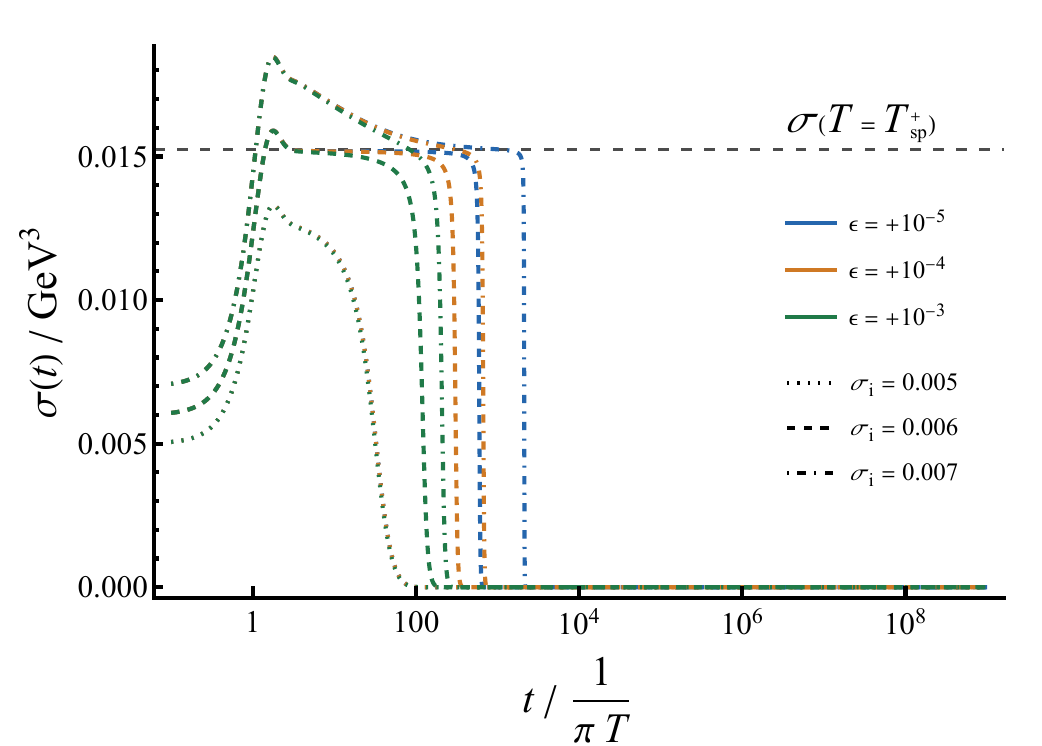}
\caption{Real-time chiral condensate evolution at \(\mu=0.6\,\mathrm{GeV}\) for temperatures just outside the spinodal interval. (a) \(T < \Tm\) for different initial condensates, showing the slow roll at small condensate values before the rapid rise toward the non-zero equilibrium. (b) \(T > \Tp\), where the trajectories exhibit an extended slow roll near the non-zero merger point before ultimately decaying to zero.}
\label{fig:first_order_condensate_evolution}
\end{figure}

Figure~\ref{fig:first_order_condensate_evolution} compares the resulting real-time trajectories. Remarkably, despite the radically different final states, both endpoints exhibit long-lived linear slow rolls. Below the lower spinodal (Fig.~\ref{fig:first_order_condensate_evolution}a), the initial increase is followed by an extended interval of slow evolution at small condensate values, before a rapid rise toward the non-zero equilibrium. The intermediate condensate depends on the initial condition, and smaller initial condensates (e.g., \(10^{-6}\,\mathrm{GeV}^3\)) lead to longer slow-roll stages. Above the upper spinodal (Fig.~\ref{fig:first_order_condensate_evolution}b), the larger initial condensates undergo a long-lived slow roll near the finite merger point before eventually decaying toward zero. The stage becomes more extended as the temperature approaches \(\Tp\) from above (i.e., as \(\epsilon\) decreases).

To confirm that this slow evolution is governed by the same geometric mechanism, we examine the bulk-field trajectories. Similar to the second-order case, the initial transients follow distinct paths, but the subsequent trajectories rapidly collapse onto a common curve at each radial position. This waveform collapse confirms that the bulk field is dynamically slaved to a slow configuration space, regardless of the endpoint structure. We therefore apply the same adiabatic reduction, showing that the distinct local geometries at the two spinodal endpoints will dictate fundamentally different drift laws.

\subsection{Reduced dynamics near the lower spinodal}\label{sec:lower}

At the lower spinodal \(\Tm\), the zero-condensate background loses local stability. Because this endpoint is still located at the symmetric point (\(\sigma=0\)), the exact \(Z_2\) chiral symmetry forbids even powers of \(\sigma\). Remarkably, the linearized operator here develops the same zero mode as at the second-order critical point, meaning the linear restoring force vanishes (\(c_1(\Tm)=0\)).

However, the local geometry differs fundamentally from \(T_c\). Our analytical projection yields a positive cubic coefficient (\(c_3 > 0\)), in stark contrast to the negative value at \(T_c\). Consequently, the cubic term now drives the condensate away from the origin, producing a slow growth rather than decay. To capture the nonlinear correction to this growth, we retain the next symmetry-allowed contribution and extend the expansion to fifth order, where the resulting \(c_5 < 0\) opposes the growth. We write:
\begin{equation}
\label{eq:lower_spinodal_fifth_order_expansion}
\begin{aligned}
\bar{\chi}(z,\sigma)
&=\sigma\chi_1(z)+\sigma^3\chi_3(z)+\sigma^5\chi_5(z)
+O(\sigma^7),\\
\mathcal{F}(\sigma)
&=c_1\sigma+c_3\sigma^3+c_5\sigma^5+O(\sigma^7).
\end{aligned}
\end{equation}

We determine \(c_1\) and \(\chi_1\) from the generalized eigenvalue problem, following the mode continuously connected to the lower-spinodal zero mode. The cubic and quintic coefficients are then extracted via the constrained boundary-value problems with \(v_{40}\) and \(v_{60}\) replaced by \(v_4(z)\) and \(v_6(z)\).

To quantify the temperature deviation from the lower spinodal, we define the relative detuning \(\epsilon = (T-\Tm)/\Tm\). For the three temperatures below \(\Tm\) (i.e., \(\epsilon < 0\)), the analytically derived coefficients are summarized in Table~\ref{tab:lower_spinodal_reduced_coefficients}. We note that \(c_1(T)\) changes sign across the endpoint, while the nonlinear coefficients remain remarkably stable.

\begin{table}[t]
\centering
\caption{Analytically derived drift coefficients near the lower spinodal, with \(\epsilon=(T-\Tm)/\Tm\). The coefficients $c_1, c_3, c_5$ are in units of \(\mathrm{GeV}\), \(\mathrm{GeV}^{-5}\), and \(\mathrm{GeV}^{-11}\), respectively.}
\label{tab:lower_spinodal_reduced_coefficients}
\renewcommand{\arraystretch}{1.15}
\begin{tabular*}{0.98\columnwidth}{@{\extracolsep{\fill}}cccc@{}}
\hline
\(\epsilon\) & \(c_1(T)\) & \(c_3(T)\) & \(c_5(T)\) \\
\hline
\(0\)        & \(0                \) & \(892.72\) & \(-4.17\times10^{6}\) \\
\(-10^{-5}\) & \(4.86\times10^{-6}\) & \(892.71\) & \(-4.17\times10^{6}\) \\
\(-10^{-4}\) & \(4.86\times10^{-5}\) & \(892.65\) & \(-4.17\times10^{6}\) \\
\(-10^{-3}\) & \(4.86\times10^{-4}\) & \(892.06\) & \(-4.16\times10^{6}\) \\
\hline
\end{tabular*}
\end{table}

\begin{figure}[h]
\centering
\includegraphics[width=0.98\columnwidth]{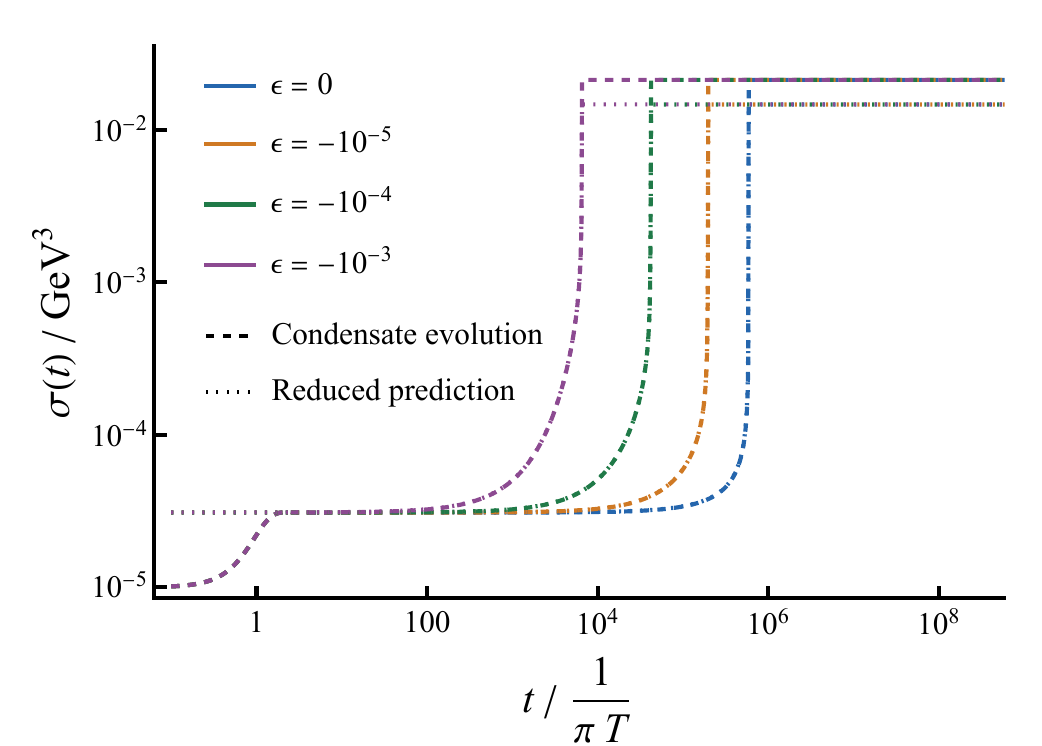}
\caption{Condensate evolution (dashed) and reduced predictions (dotted) near the lower spinodal at $\mu=0.6\,\mathrm{GeV}$ for $\sigma_i=10^{-5}\,\mathrm{GeV}^3$. The temperature deviations are $\epsilon=0,\,-10^{-5},\,-10^{-4},\,-10^{-3}$, with $T=\Tm(1+\epsilon)$. The predictions are obtained by integrating Eq.~\eqref{eq:lower_spinodal_reduced_dynamics}, using the coefficients in Table~\ref{tab:lower_spinodal_reduced_coefficients}. Each prediction is initialized at $t_0\simeq2/(\pi T)$ with $\sigma_0=\sigma(t_0)$ taken from the full numerical solution.}
\label{fig:first_order_lower_prediction}
\end{figure}

We numerically integrate the fifth-order reduced equation
\begin{equation}
\label{eq:lower_spinodal_reduced_dynamics}
\frac{d\sigma}{dt}
=c_1(T)\sigma+c_3(T)\sigma^3+c_5(T)\sigma^5,
\end{equation}
using \(\sigma_0=\sigma(t_0)\) from the full numerical evolution at \(t_0 \simeq 2/(\pi T)\). Figure~\ref{fig:first_order_lower_prediction} compares the resulting predictions with the bulk evolution for \(\sigma_i=10^{-5}\,\mathrm{GeV}^3\). The reduced predictions closely follow the long-lived slow-roll stage and accurately reproduce its increasing duration as the temperature approaches \(\Tm\) from below. As the condensate leaves the slow-roll region and approaches the non-zero equilibrium state, the fifth-order expansion (which is local about \(\sigma=0\)) begins to deviate, indicating the onset of higher-order corrections.

\subsection{Reduced dynamics near the upper spinodal}\label{sec:upper}

We next consider the slow evolution above the upper spinodal at \(\mu = 0.6\,\mathrm{GeV}\). Unlike the lower spinodal, the upper endpoint \(\Tp\) is characterized by the merger of a non-zero minimum with the intervening barrier. The slow roll therefore occurs around a finite background condensate \(\sigma_{\mathrm{sp}}^{+} = \sigma_{\mathrm{eq}}(\Tp) = 0.015246129\,\mathrm{GeV}^3\) ($\sigma_{\mathrm{eq}}(T)$ denotes the equilibrium value of the condensate at temperature $T$). Although this stationary branch no longer exists for \(T > \Tp\), the nearby dynamics remains slow as trajectories pass through the vicinity of the endpoint profile.

We define the condensate displacement relative to the endpoint as:
\begin{equation}
\label{eq:upper_spinodal_variables}
\delta \sigma(t)=\sigma(t)-\sigma_{\mathrm{sp}}^{+},
\qquad\epsilon=\frac{T-T_{\mathrm{sp}}^+}{T_{\mathrm{sp}}^+}.
\end{equation}
Because the expansion is centered on a non-zero background, the \(Z_2\) symmetry is absent, and even powers of \(\delta \sigma\) are allowed. The local field profile and the reduced drift are expanded as:
\begin{equation}
\label{eq:upper_spinodal_expansion}
\begin{aligned}
\bar{\chi}(z,\delta \sigma,\epsilon)
&=\chi_{\mathrm{sp}}^{+}(z)+\delta \sigma\chi_1(z)
+\epsilon\chi_\epsilon(z)+\delta \sigma^2\chi_2(z)+\cdots,\\
\frac{d\delta \sigma}{dt}
&=a\epsilon+b \delta \sigma^2+O(\epsilon \delta \sigma,\delta \sigma^3,\epsilon^2).
\end{aligned}
\end{equation}
The constant drift proportional to \(\epsilon\) accounts for the fact that this endpoint profile is no longer stationary when the temperature is shifted away from \(\Tp\).

At the upper spinodal $T=\Tp$, the reference profile $\chi_{\mathrm{sp}}^{+}(z)$ is a nonzero static solution satisfying $R[\chi_{\mathrm{sp}}^{+}]=0$, with boundary condensate $\sigma_{\mathrm{sp}}^{+}$. The coefficients $a$ and $b$ are determined from the constrained radial problems described in Appendix~\ref{app:finite_temperature_reduction}. The resulting leading-order reduced equation is
\begin{equation}
\label{eq:upper_spinodal_reduced_dynamics}
\frac{d\sigma}{dt}
=a\epsilon+b\left[\sigma-\sigma_{\mathrm{sp}}^{+}\right]^2.
\end{equation}

\begin{figure}[tbp]
\centering
\includegraphics[width=0.98\columnwidth]{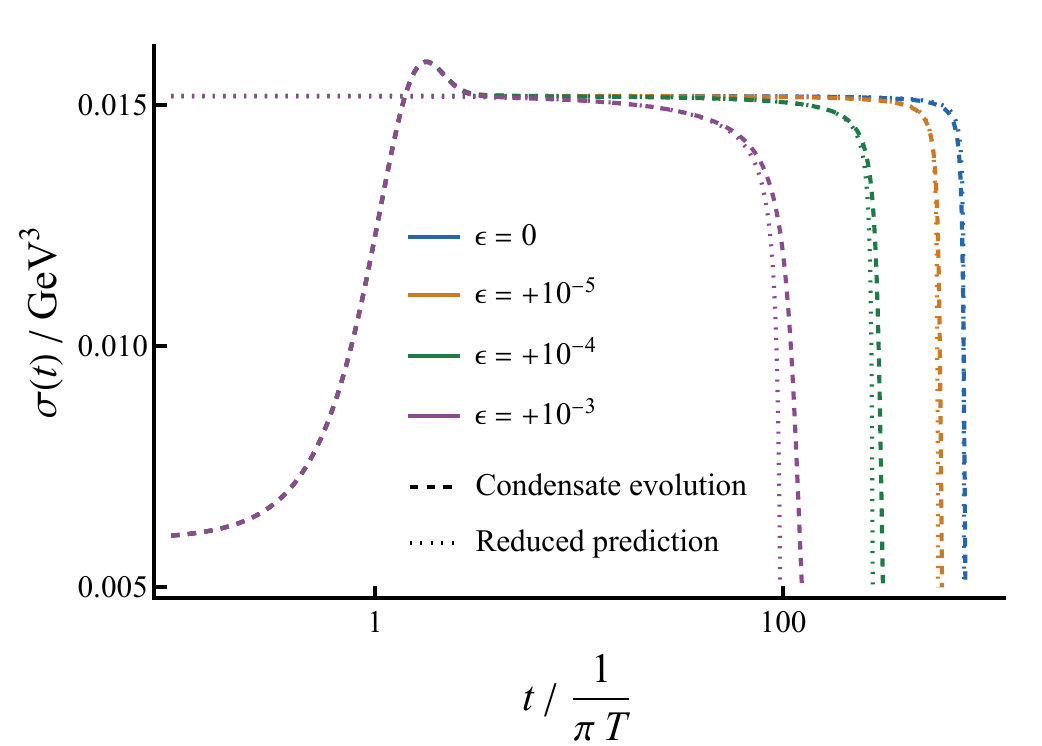}
\caption{Condensate evolution (dashed) and reduced predictions (dotted) near the upper spinodal at $\mu=0.6\,\mathrm{GeV}$ for $\sigma_i=0.007\,\mathrm{GeV}^3$. The temperature deviations are $\epsilon=0,\,+10^{-5},\,+10^{-4},\,+10^{-3}$, with $T=\Tp(1+\epsilon)$. The predictions are obtained by integrating Eq.~\eqref{eq:upper_spinodal_reduced_dynamics}, using the coefficients in Eq.~\eqref{eq:upper_spinodal_coefficient_values}. Each prediction is initialized at $t_0\simeq2/(\pi T)$ with $\sigma_0=\sigma(t_0)$ taken from the full numerical solution.}
\label{fig:upper_spinodal_reduced_prediction}
\end{figure}

Near the endpoint condensate, the quadratic contribution is small and the drift is approximately constant, producing the nearly linear slow-roll evolution. As the temperature approaches \(\Tp\) from above, the temperature-induced drift decreases, allowing trajectories to remain near the endpoint profile for longer times. For our model parameters, we obtain
\begin{equation}
\label{eq:upper_spinodal_coefficient_values}
a\simeq-0.0095\,\mathrm{GeV}^{4},
\qquad
b\simeq-22.90\,\mathrm{GeV}^{-2},
\end{equation}
 We integrate Eq.~\eqref{eq:upper_spinodal_reduced_dynamics} from \(t_0 \simeq 2/(\pi T)\), taking \(\sigma(t_0)\) from the full numerical evolution. Figure~\ref{fig:upper_spinodal_reduced_prediction} compares the resulting predictions with trajectories initialized at \(\sigma_i=0.007\,\mathrm{GeV}^3\) for $\epsilon=0$, $+10^{-5}$, $+10^{-4}$, and $+10^{-3}$.

The reduced predictions closely follow the numerical trajectories during the long-lived slow-roll stage and capture its increasing duration as the temperature approaches \(\Tp\) from above. This agreement supports the local description in terms of the temperature-induced drift and the quadratic dependence on the displacement from the endpoint condensate.

\section{Effective Potential Landscape and the Longevity of Slow Roll}
\label{sec:landscape}
\subsection{Effective potential geometry and slow-roll duration}
\label{sec:potential_geometry_duration}
The reduced evolution equations derived above connect the bulk dynamics to a time-dependent Ginzburg--Landau description of the condensate. Absorbing the condensate-independent kinetic coefficient into the potential, we write $U_{\mathrm{eff}}=\Gamma V_{\mathrm{eff}}$ \footnote{It is important to distinguish this dynamical effective potential \(U_{\mathrm{eff}}(\sigma,T)\) from the static thermodynamic grand potential \(\mathcal{F}\) computed in Sec.~\ref{sec:model}. The latter is derived from the Euclidean on-shell action and governs the equilibrium phase diagram. In contrast, \(U_{\mathrm{eff}}\) is defined by integrating the real-time drift, \(\mathcal{F}(\sigma) = -\partial_\sigma U_{\mathrm{eff}}\), and its profile inherently encodes the non-equilibrium dissipative dynamics. While the local extrema of \(U_{\mathrm{eff}}\) must coincide with those of the thermodynamic potential to ensure the correct equilibrium states, its curvature dictates the condensate's acceleration during the intermediate stage. This distinction is precisely what allows the dynamical landscape to exhibit the ``double-zero'' flatness responsible for the long-lived linear slow roll, even when the system is far from equilibrium.} with $\Gamma>0$, so that
\begin{equation}
\label{eq:GL_with_mobility}
\dot\sigma=-\partial_\sigma U_{\mathrm{eff}}(\sigma,T).
\end{equation}
For the expansions about the zero-condensate background, the reduced drift and its associated potential are
\begin{equation}
\label{eq:HGL_fifth_order_drift}
\dot\sigma=c_1(T)\sigma+c_3(T)\sigma^3+c_5(T)\sigma^5,
\end{equation}
\begin{equation}
\label{eq:HGL_fifth_order_potential}
U_{\mathrm{eff}}(\sigma,T)=-\frac{c_1(T)}{2}\sigma^2-\frac{c_3(T)}{4}\sigma^4-\frac{c_5(T)}{6}\sigma^6.
\end{equation}
The coefficients are supplied by the bulk reduction in Tables~\ref{tab:second_order_reduced_coefficients} and~\ref{tab:lower_spinodal_reduced_coefficients}. Near $T_c$, $c_3<0$, and the quartic potential gives the leading nonlinear restoring force. Near $T_{\mathrm{sp}}^-$, $c_3>0$ and $c_5<0$, so the quartic term drives growth and the sextic term suppresses it at larger condensates. The small-condensate potential geometry at both points is determined by the quadratic and quartic terms.

At the upper spinodal, the expansion is instead centered on the nonzero endpoint condensate $\sigma_{\mathrm{sp}}^+=\sigma_{\mathrm{eq}}(T_{\mathrm{sp}}^+)$. Assuming the sixth-order holographic Ginzburg–Landau (HGL) form and matching its endpoint expansion to the quadratic coefficient $b$ from Eq.~\eqref{eq:upper_spinodal_coefficient_values}, the evolution equation through fifth order is
\begin{equation}
\label{eq:upper_detuned_fifth_drift}
\begin{split}
\dot\sigma={}&a\epsilon+b(\sigma-\sigma_{\mathrm{sp}}^+)^2+\frac{2b}{\sigma_{\mathrm{sp}}^+}(\sigma-\sigma_{\mathrm{sp}}^+)^3\\
&+\frac{5b}{4(\sigma_{\mathrm{sp}}^+)^2}(\sigma-\sigma_{\mathrm{sp}}^+)^4+\frac{b}{4(\sigma_{\mathrm{sp}}^+)^3}(\sigma-\sigma_{\mathrm{sp}}^+)^5.
\end{split}
\end{equation}
Here $a$ and $b$ are fixed by the upper-spinodal reduction. Throughout this section, $T=T_c(1+\epsilon)$ or $T=T_{\mathrm{sp}}^\pm(1+\epsilon)$ for the point under consideration.
\begin{figure*}[t]
\centering
\subfloat[Near $T_c$\label{fig:potential_second}]{\includegraphics[width=0.32\textwidth]{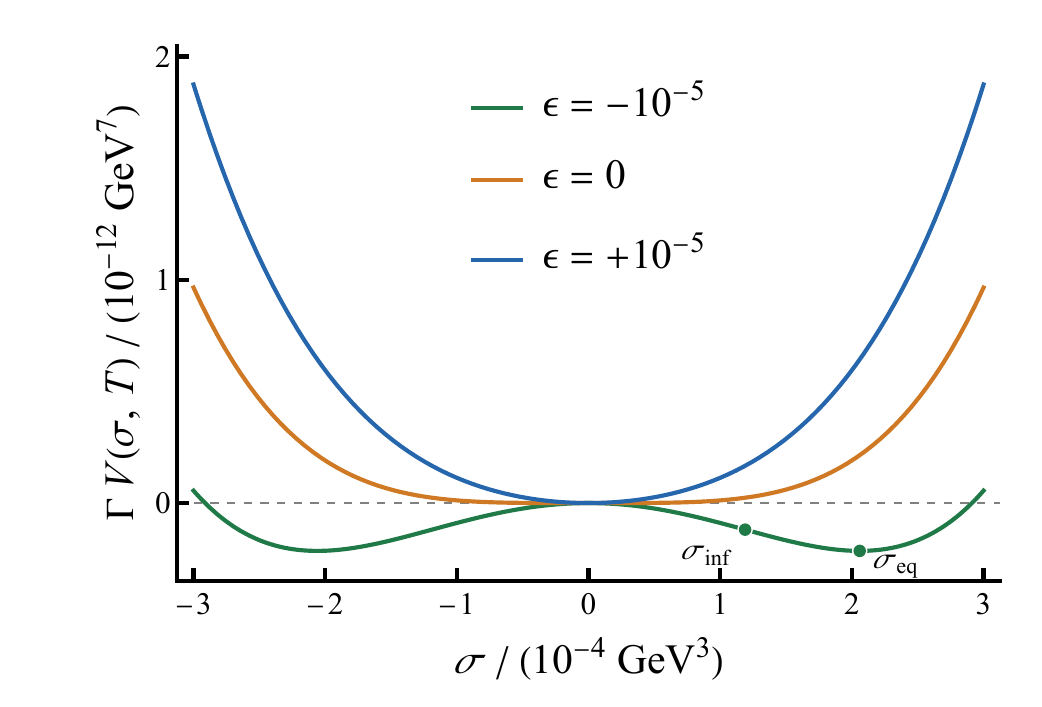}}\hfill
\subfloat[Near $T_{\mathrm{sp}}^-$\label{fig:potential_lower}]{\includegraphics[width=0.32\textwidth]{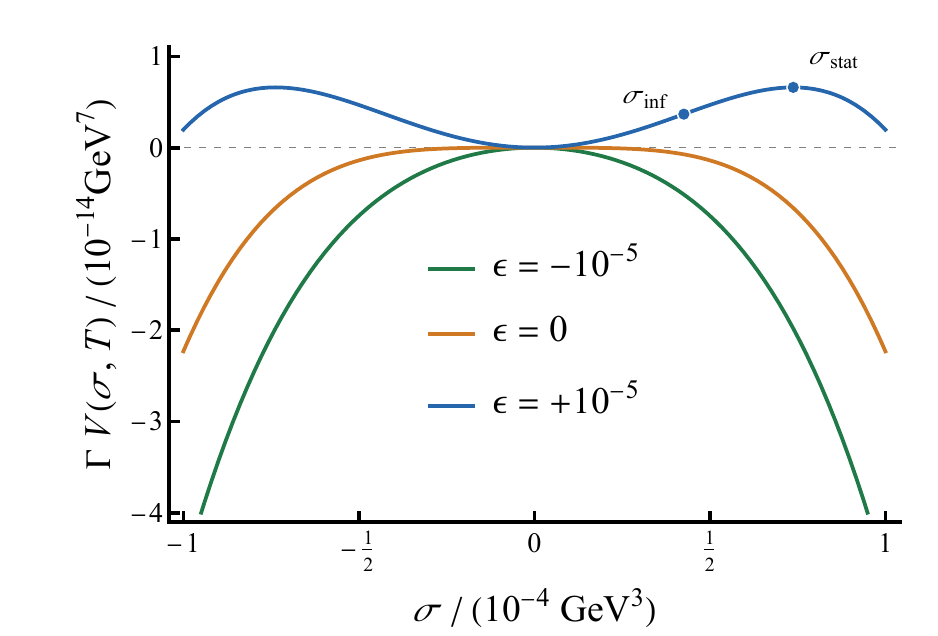}}\hfill
\subfloat[Near $T_{\mathrm{sp}}^+$\label{fig:potential_upper}]{\includegraphics[width=0.32\textwidth]{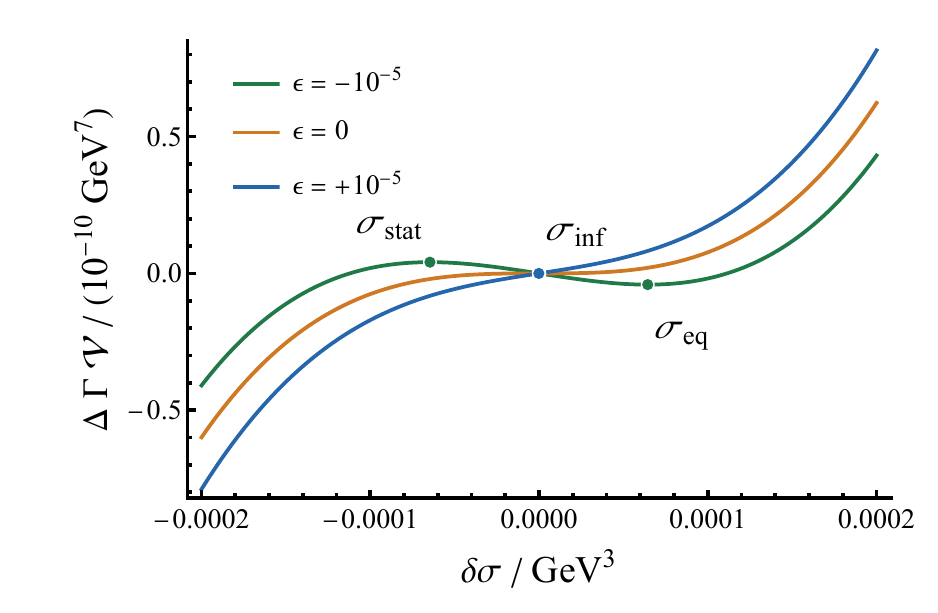}}
\caption{Local effective-potential landscapes near the three characteristic points. Green, orange, and blue curves correspond to $\epsilon=-10^{-5},0,+10^{-5}$, respectively. Panels~(a) and~(b) show the local potentials near zero condensate. Panel~(c) shows the local potential about $\sigma_{\mathrm{sp}}^+$, with its value at that condensate subtracted. The marked points identify potential inflections and stationary points.}
\label{fig:potential_landscapes}
\end{figure*}

Figure~\ref{fig:potential_landscapes}(a) shows the potential near the second-order transition. At $T_c$, the quadratic term vanishes and the origin is a flat quartic minimum. Below $T_c$, the origin becomes a local maximum between two nonzero minima, while above $T_c$ it is the only minimum. On the positive-condensate branch, evolution therefore approaches a nonzero equilibrium below $T_c$ and the symmetric state above it. In these figures, $\sigma_{\mathrm{inf}}$ denotes an inflection point of the effective potential, $\sigma_{\mathrm{eq}}$ a local minimum corresponding to a stable or metastable equilibrium, and $\sigma_{\mathrm{stat}}$ the unstable stationary point at the barrier maximum.

The lower-spinodal landscape in Fig.~\ref{fig:potential_landscapes}(b) has the opposite quartic sign. At $T_{\mathrm{sp}}^-$, the origin is a flat local maximum, and small positive condensates grow away from it. Below the endpoint, the quadratic term reinforces this growth. Above it, the origin becomes a local minimum separated from larger condensates by a barrier. The nonzero inflection point lies between the origin and the barrier.

Near the upper spinodal, Fig.~\ref{fig:potential_landscapes}(c) shows the potential difference
\begin{equation}
\label{eq:upper_potential_difference}
\Delta U_{\mathrm{eff}}(\sigma,T)\equiv U_{\mathrm{eff}}(\sigma,T)-U_{\mathrm{eff}}(\sigma_{\mathrm{sp}}^+,T)
\end{equation}
A local maximum and minimum near $\sigma_{\mathrm{sp}}^+$ exist below $T_{\mathrm{sp}}^+$ and merge into a stationary inflection point at the endpoint. Above it, these stationary points disappear, leaving a region of weak potential gradient through which the condensate decays slowly.

These landscapes distinguish a small evolution rate from a slowly varying evolution rate. At fixed temperature, the former is controlled by the potential gradient in Eq.~\eqref{eq:GL_with_mobility}, the latter follows from
\begin{equation}
\label{eq:near_critical_acceleration}
\ddot\sigma=-U_{\mathrm{eff}}''(\sigma,T)\dot\sigma,
\end{equation}
where primes denote condensate derivatives. A small curvature suppresses the relative change of the drift and allows an approximately linear trajectory to persist. We estimate its duration by expanding around the entrance time $t_L$ through $(t-t_L)^3$ and applying the same linear-fit threshold $R_{\mathrm{th}}^2=0.99999$ used to identify the numerical intervals. Appendix~\ref{app:duration_derivation} gives the resulting criterion. The comparisons below retain the cubic time correction, which also determines the duration near a nonstationary inflection point where the quadratic correction vanishes.

\begin{figure*}[t]
\centering
\subfloat[$\epsilon=-10^{-5}$\label{fig:duration_Tc_below}]{\includegraphics[width=0.32\textwidth]{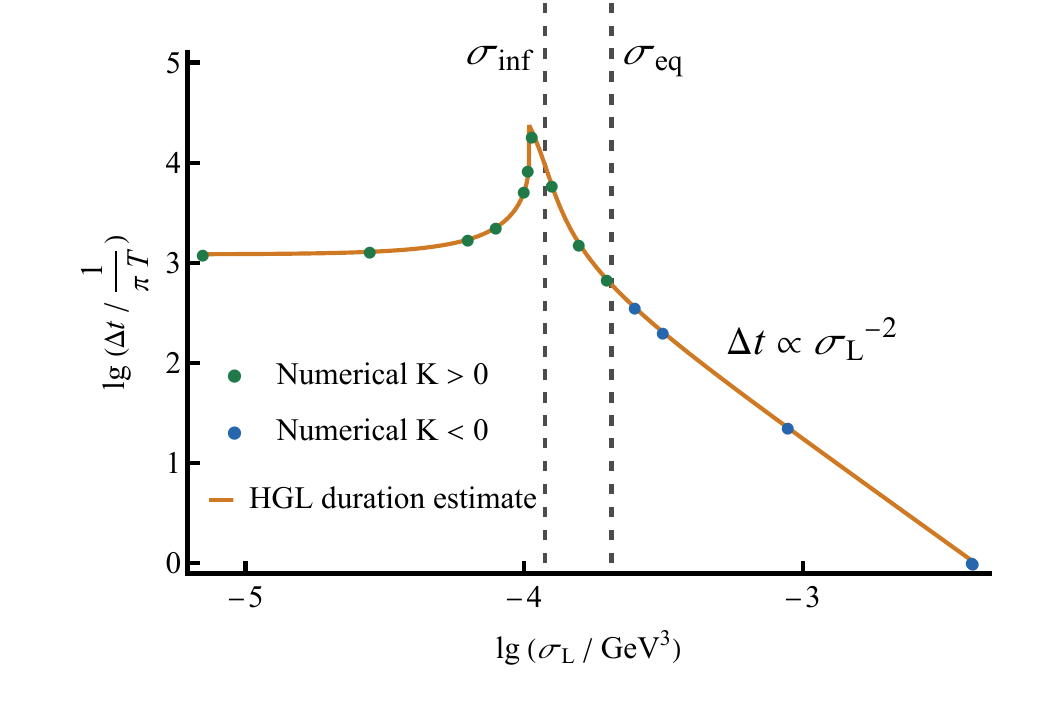}}\hfill
\subfloat[$\epsilon=0$\label{fig:duration_Tc_at}]{\includegraphics[width=0.32\textwidth]{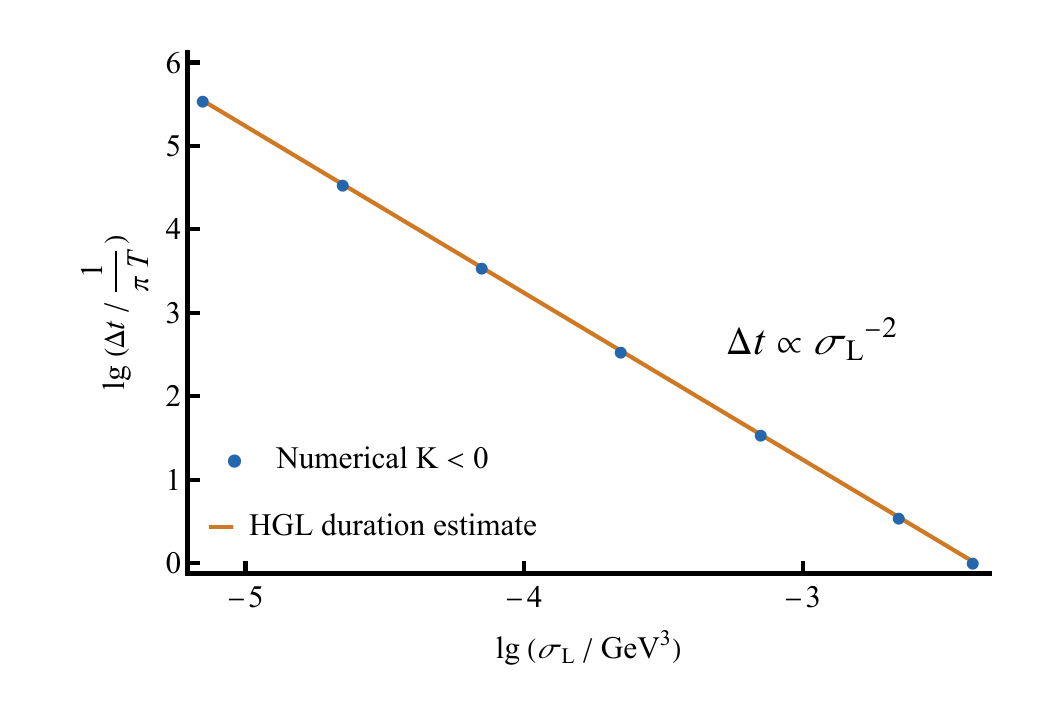}}\hfill
\subfloat[$\epsilon=+10^{-5}$\label{fig:duration_Tc_above}]{\includegraphics[width=0.32\textwidth]{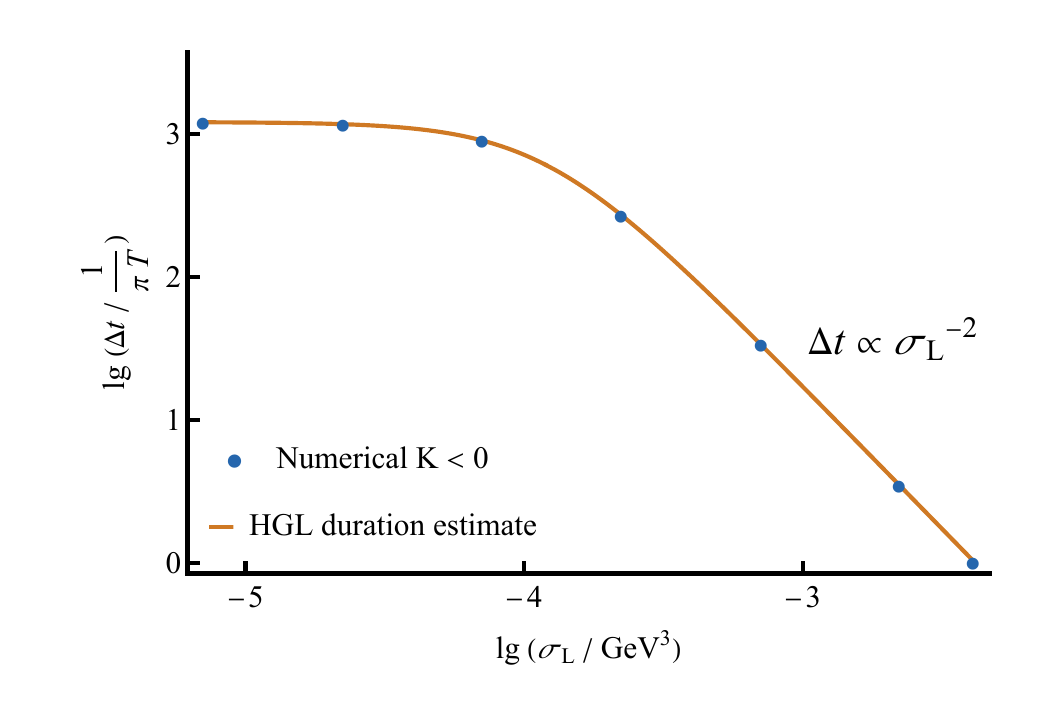}}
\caption{Slow-roll duration near $T_c$, with $T=T_c(1+\epsilon)$. Points show the numerical intervals, with green and blue indicating positive and negative slopes, respectively. Orange curves show the duration estimates obtained from Eqs.~\eqref{eq:duration_cubic_r_squared} and~\eqref{eq:duration_cubic_threshold}, using the reduced coefficients in Table~\ref{tab:second_order_reduced_coefficients}. The dashed lines in panel~(a) mark the positive inflection point and equilibrium minimum.}
\label{fig:duration_second_at}
\label{fig:duration_detuned}
\end{figure*}

We first consider the three temperatures near $T_c$ in Fig.~\ref{fig:duration_second_at}.  At $T_c$, the quartic minimum gives $U_{\mathrm{eff}}''=-3c_3(T_c)\sigma_L^2$, producing the inverse-square dependence
\begin{equation}
\label{eq:critical_scaling}
\Delta t\propto\sigma_L^{-2}.
\end{equation}

Below $T_c$, the quadratic and quartic contributions to the curvature cancel at $\sigma_{\mathrm{inf}}$, producing the enhanced duration in Fig.~\ref{fig:duration_second_at}(a). The condensate is still evolving at the inflection point, but its rate changes weakly, whereas at the minimum the driving force itself vanishes. Above $T_c$, both contributions to the curvature are positive. The duration then approaches a constant at small $\sigma_L$, where the quadratic potential dominates, and crosses over to inverse-square behavior as the quartic contribution increases. The three panels connect the duration directly to the changing potential geometry in Fig.~\ref{fig:potential_landscapes}(a).

\begin{figure*}[t]
\centering
\subfloat[$\epsilon=-10^{-5}$\label{fig:duration_lower_below}]{\includegraphics[width=0.32\textwidth]{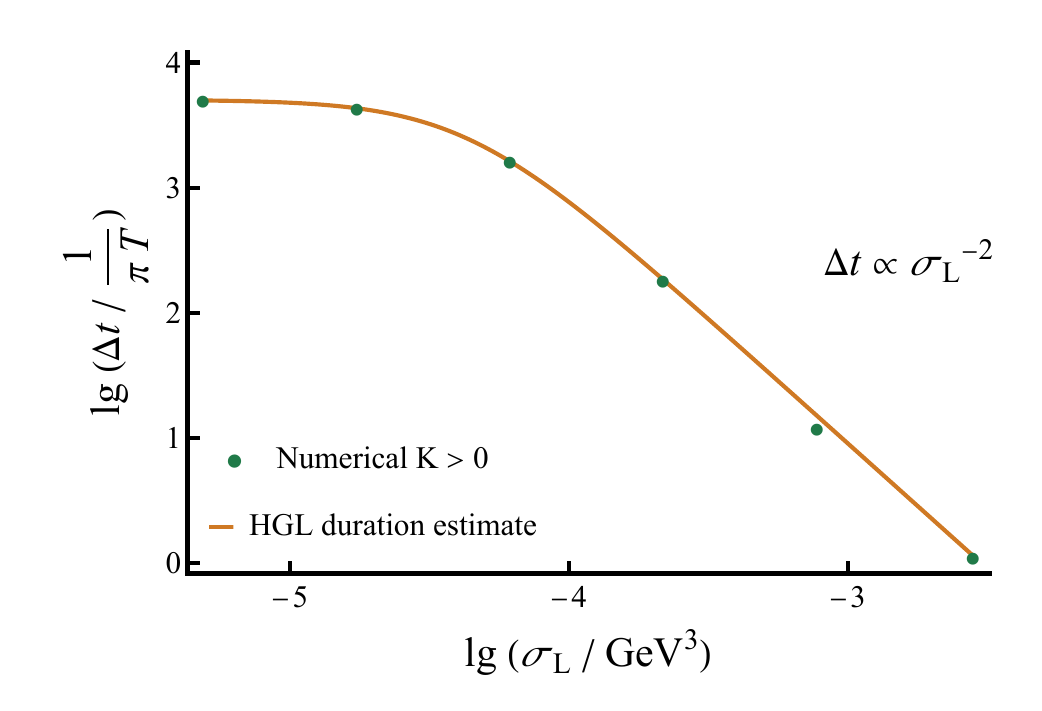}}\hfill
\subfloat[$\epsilon=0$\label{fig:duration_lower_at}]{\includegraphics[width=0.32\textwidth]{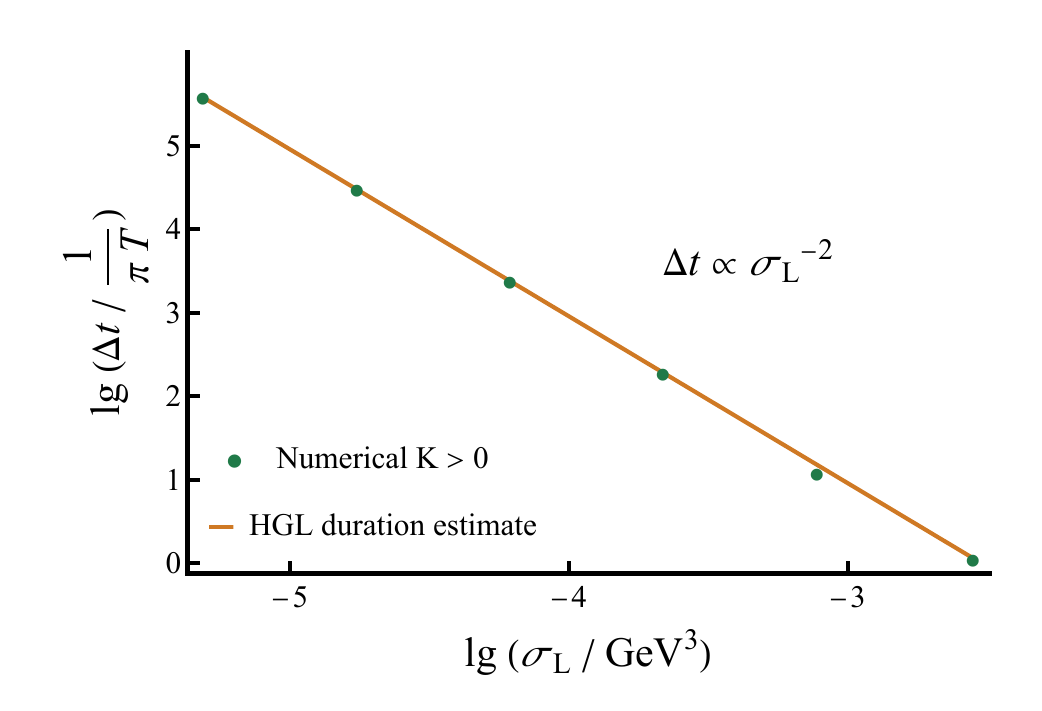}}\hfill
\subfloat[$\epsilon=+10^{-5}$\label{fig:duration_lower_above}]{\includegraphics[width=0.32\textwidth]{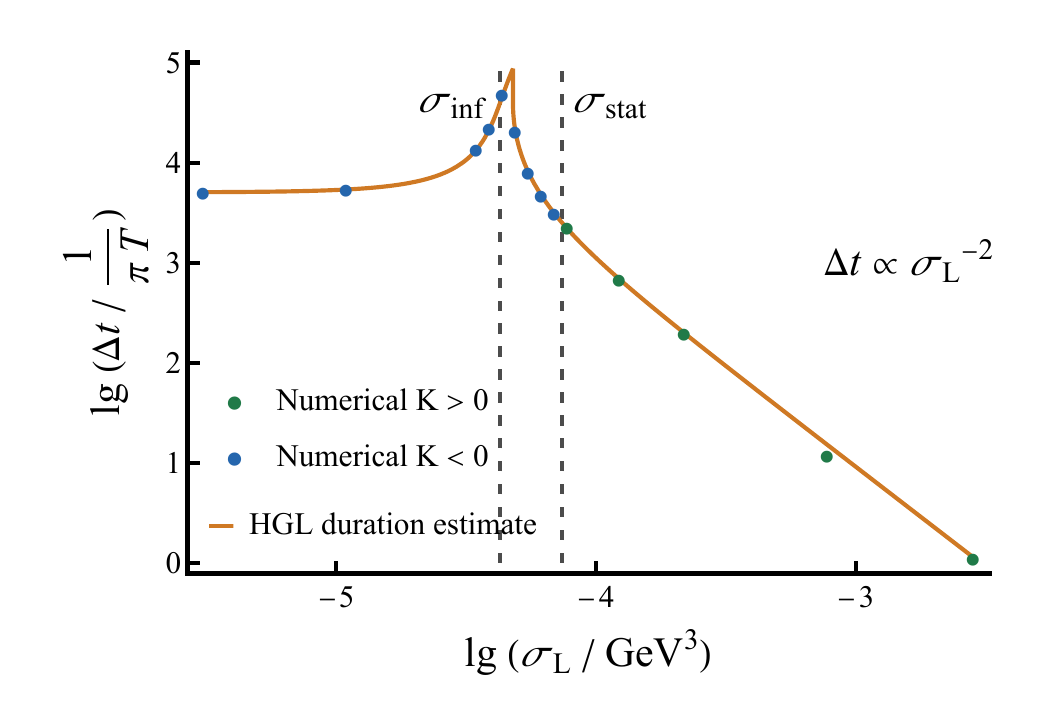}}
\caption{Slow-roll duration near $T_{\mathrm{sp}}^-$, with $T=T_{\mathrm{sp}}^-(1+\epsilon)$. Green and blue points indicate positive and negative numerical slopes. Orange curves show the duration estimates obtained from Eqs.~\eqref{eq:duration_cubic_r_squared} and~\eqref{eq:duration_cubic_threshold}, using the near-endpoint coefficients from Table~\ref{tab:lower_spinodal_reduced_coefficients}. The dashed lines in panel~(c) mark the positive inflection point and the barrier maximum.}
\label{fig:duration_lower}
\end{figure*}

Figure~\ref{fig:duration_lower} presents the corresponding comparison near $T_{\mathrm{sp}}^-$. At the endpoint, the negative quartic potential again gives $|U_{\mathrm{eff}}''|\propto\sigma_L^2$, and panel~(b) exhibits $\Delta t\propto\sigma_L^{-2}$. Thus, the same duration scaling accompanies condensate growth at $T_{\mathrm{sp}}^-$ and condensate decay at $T_c$. It reflects the magnitude and condensate dependence of the curvature, despite the opposite directions of motion.

Below $T_{\mathrm{sp}}^-$, the quadratic and quartic terms reinforce one another in the curvature. Panel~(a) consequently shows a small-condensate duration plateau followed by inverse-square behavior. Above the endpoint, the positive quadratic curvature competes with the negative quartic contribution, producing a nonzero inflection point and the duration peak in panel~(c). The barrier maximum lies at a larger condensate and separates decay toward the origin from growth toward the broken phase. As near $T_c$, the duration enhancement is associated with the inflection point rather than the stationary point. The side of the endpoint on which this enhancement occurs is reversed because the quartic coefficient has the opposite sign.

\begin{figure*}[t]
\centering
\subfloat[$\epsilon=-10^{-5}$\label{fig:duration_upper_below}]{\includegraphics[width=0.32\textwidth]{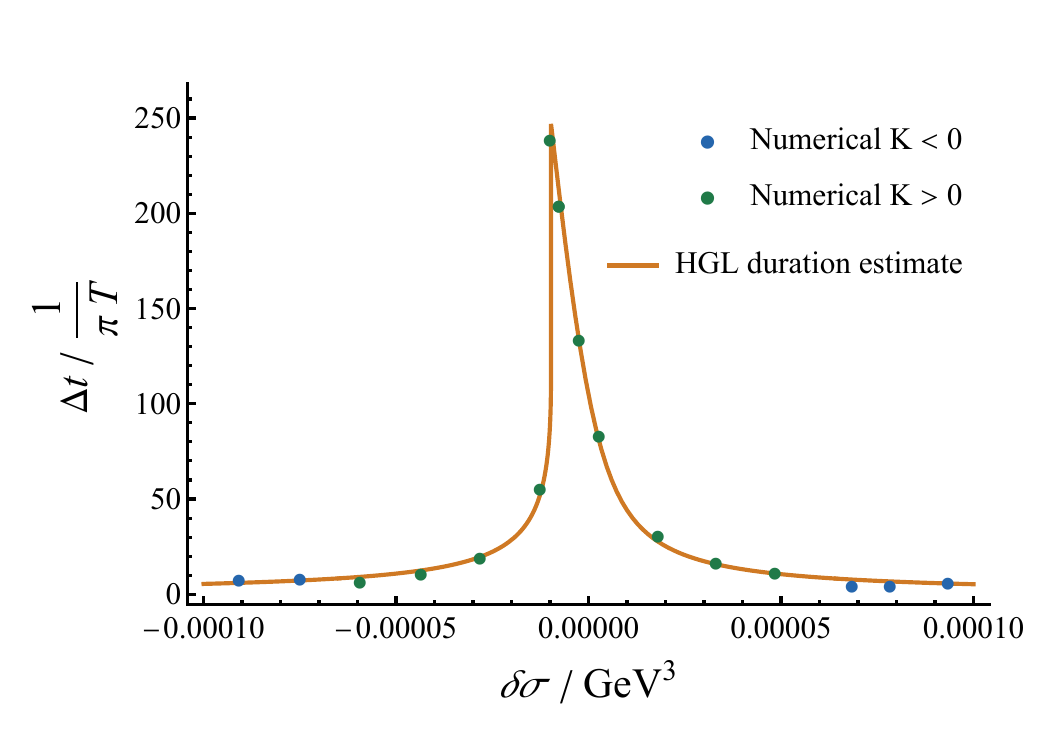}}\hfill
\subfloat[$\epsilon=0$\label{fig:duration_upper_at}]{\includegraphics[width=0.32\textwidth]{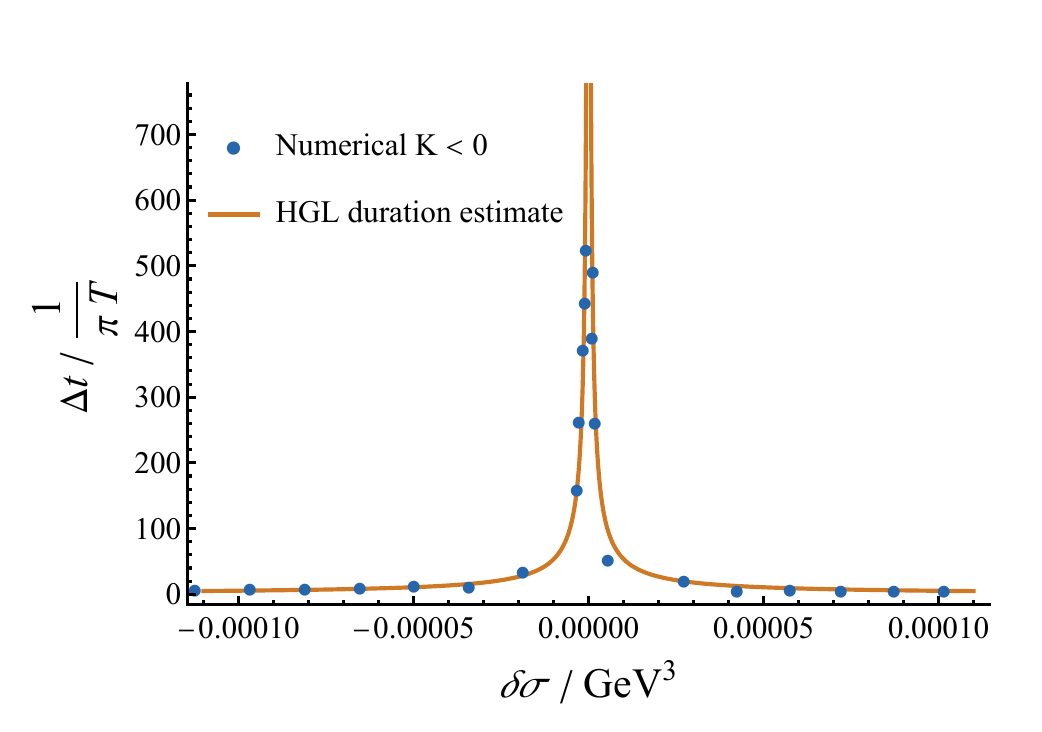}}\hfill
\subfloat[$\epsilon=+10^{-5}$\label{fig:duration_upper_above}]{\includegraphics[width=0.32\textwidth]{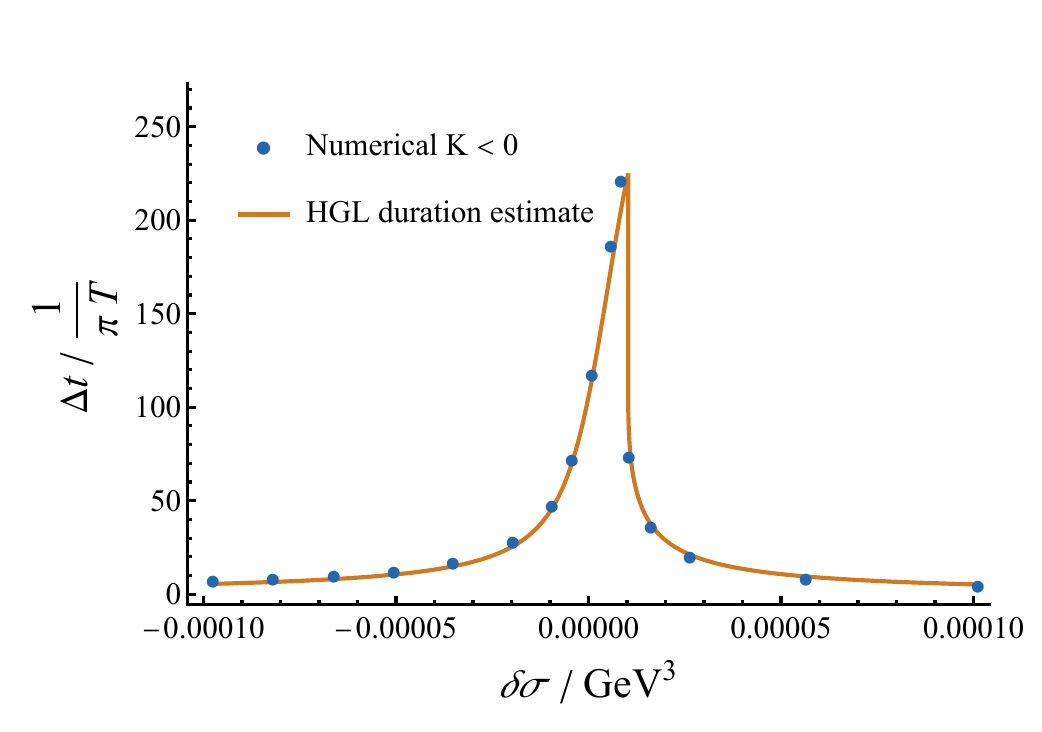}}
\caption{Slow-roll duration near $T_{\mathrm{sp}}^+$, with $T=T_{\mathrm{sp}}^+(1+\epsilon)$ and $\delta\sigma=\sigma_L-\sigma_{\mathrm{sp}}^+$. Green and blue points indicate positive and negative numerical slopes. Orange curves show the duration estimates obtained from Eqs.~\eqref{eq:duration_cubic_r_squared} and~\eqref{eq:duration_cubic_threshold}.}
\label{fig:upper_durations}
\end{figure*}

Finally, Fig.~\ref{fig:upper_durations} shows the durations near $T_{\mathrm{sp}}^+$ as functions of the entrance displacement $\delta\sigma=\sigma_L-\sigma_{\mathrm{sp}}^+$. The predictions use the full fifth-order equation~\eqref{eq:upper_detuned_fifth_drift}. At the endpoint, its leading quadratic drift gives $U_{\mathrm{eff}}''\simeq-2b\delta\sigma$, leading to
\begin{equation}
\label{eq:upper_duration_scaling}
\Delta t\propto|\delta\sigma|^{-1}.
\end{equation}
The concentration of long intervals near $\delta\sigma=0$ reflects this linear vanishing of the curvature, in contrast to its quadratic vanishing at the zero-condensate endpoints.

Below $T_{\mathrm{sp}}^+$, the enhanced durations occur between the nearby maximum and minimum of the potential, where trajectories can evolve slowly toward the nonzero equilibrium. At the endpoint, these stationary points merge and the slow region is centered on the nonzero endpoint condensate. Above $T_{\mathrm{sp}}^+$, the duration enhancement persists while the condensate decays through a region without a local equilibrium. Temperature detuning changes the potential gradient and hence the cubic time correction, even though the retained term $a\epsilon$ does not contribute directly to $U_{\mathrm{eff}}''$.

The three comparisons identify a common origin of long-lived linear slow roll. A weak gradient produces slow motion, while a small curvature maintains an almost constant rate. The quartic geometry at \(T_c\) and \(T_{\mathrm{sp}}^-\) yields an inverse-square duration scaling, whereas the cubic local geometry at \(T_{\mathrm{sp}}^+\) yields an inverse-displacement scaling. Consequently, for a fixed displacement $|\delta\sigma|$, the curvature near $T_{\mathrm{sp}}^+$ vanishes only linearly,
$|U_{\mathrm{eff}}''|\simeq 2|b|\,|\delta\sigma|$, whereas at the symmetric endpoints it vanishes quadratically, $|U_{\mathrm{eff}}''|\simeq 3|c_3|\sigma^2$. At equal displacement the former is therefore larger and the linear window is shorter; the long durations found near $T_{\mathrm{sp}}^+$ at $\epsilon=0$ originate instead from
the trajectories approaching $\delta\sigma=0$ asymptotically. Crucially, it is worth emphasizing that these scaling laws hold over a remarkably broad range of the entrance condensate, extending up to a few fractions of the vacuum condensate (of order \(10^{-2}\,\mathrm{GeV}^3\)). This robustness implies that the long-lived linear slow roll is a generic feature for a wide class of physically relevant initial states characterized by this scale. As a result, the corresponding duration of this intermediate stage can easily reach parametrically large time scales of \(\mathcal{O}(10^3 - 10^4)/(\pi T)\), providing an exceptionally wide temporal window for the prethermal dynamics. Away from the endpoint temperatures, the inflection points further organize and extend the durations even when the condensate is not stationary.

The results above reveal that the long-lived linear slow roll is a generic consequence of the structural evolution of the effective potential near phase-transition endpoints, rather than a fine-tuned artifact. 

At the symmetric endpoints (\(T_c\) and \(\Tm\)), the exact \(Z_2\) chiral symmetry enforces a stationary point at the origin. Away from the endpoint, an inflection point \(\sigma_{\mathrm{inf}}\) emerges and the potential acquires a nonzero curvature. As \(T \to T_{\mathrm{endpoint}}\), the linear coefficient \(c_1(T)\) is driven to zero, causing the inflection point to move steadily toward the origin and collide with the stationary point. This structural merger realizes the ``double-zero'' condition \(U_{\mathrm{eff}}'(0) = U_{\mathrm{eff}}''(0) = 0\), eliminating the linear restoring force and making the cubic drift mathematically inevitable. 

At the upper spinodal \(\Tp\), the system also satisfies the ``double-zero'' condition, but at a non-zero condensate \(\sigma_{\mathrm{sp}}^+\) due to the merger of a minimum and the intervening barrier. Because the expansion is now centered on a symmetry-broken background, the local potential takes the form \(U_{\mathrm{eff}} \sim \delta \sigma ^3\), inevitably generating a quadratic drift. Therefore, regardless of whether the double-zero condition occurs at the symmetric origin or at a non-zero merger point, it universally dictates the same physics: the suppression of linear restoring forces and the emergence of long-lived slow roll.

This structural restructuring is explicitly visualized by comparing the three potential landscapes in Figs.~\ref{fig:potential_landscapes}(a), \ref{fig:potential_landscapes}(b), and~\ref{fig:potential_landscapes}(c). The direct collision of stationary points and the movement of inflection points provide a transparent geometric understanding of the distinct drift laws. Because this mechanism relies solely on the structural merging of stationary points dictated by the phase-transition structure, the resulting slow-roll behavior appears robustly across a broad class of initial conditions and physical parameters. It is intrinsically tied to the geometric nature of the phase-transition endpoints, rather than to the specific details of the holographic model considered here.

\section{Discussion and conclusions}\label{sec:conclusions}

We have systematically investigated the homogeneous chiral-condensate dynamics near the second-order transition and both first-order spinodal endpoints in a finite-density soft-wall model. A central result of our analysis is the systematic derivation of the effective boundary dynamics. By demonstrating that the bulk field is adiabatically slaved to a one-dimensional slow configuration space (invariant manifold), we successfully projected the 5D real-time partial differential equation onto a first-order Ginzburg-Landau equation, where the effective potential is derived self-consistently from the holographic model rather than treated as a phenomenological input. This effective dynamics allows us to resolve the fine structure of the previously identified intermediate stage. While it appears as a quasi-static prethermalization plateau~\cite{Cao:2023,Zheng:2025} due to the extremely small drift slope, a closer inspection of the evolution reveals that the system is intrinsically undergoing a long-lived linear slow roll tracking the quasi-steady value.

The origin and longevity of this linear slow roll are deeply rooted in the ``double-zero'' condition (\(V'=0\) and \(V''=0\)) at the phase-transition endpoints. In the holographic bulk, the vanishing of the gradient is mathematically manifested as the appearance of a marginal zero eigenvalue in the linearized radial operator, which eliminates the linear drift coefficient. Together with the vanishing curvature, this geometric flatness inherently suppresses both the drift velocity and its acceleration. This highlights a crucial distinction from conventional critical slowing down: the long-lived dynamics are not merely governed by the divergence of a relaxation time at a single critical point, but are structurally protected by the local geometric structure of the effective potential. Consequently, the distinct local structures imprint fundamentally different drift laws. At the symmetric second-order point \(T_c\), the vanishing of the linearized eigenvalue dictates that the linear drift coefficient is zero, yielding a purely nonlinear relaxation governed by the leading cubic term, with an analytically derived coefficient \(c_3(T_c) \simeq -476.58\,\mathrm{GeV}^{-5}\). At the lower spinodal \(T_{\rm sp}^-\), the same marginal condition applies, but the quartic potential coefficient changes sign, driving the condensate to grow away from the symmetric origin.

In sharp contrast, at the upper spinodal \(T_{\rm sp}^+\), the local effective potential dictates a distinct nonlinear drift governed by the quadratic term around this nonzero condensate. We further examined the effect of temperature detuning at this endpoint: a small temperature shift splits the minimum and the barrier below \(T_{\rm sp}^+\), and removes them above it. These comparisons connect the observed drift signs and amplitude dependence to the local effective potential, complementing earlier studies of late-time holographic spinodal slowing~\cite{Caddeo2024Spinodal}, with an explicit resolution of the intermediate-time linear slow roll.

Furthermore, the local potential curvature provides a robust estimate of the slow-roll interval duration, yielding an inverse-square amplitude dependence at the second-order critical point and lower spinodal, and an inverse-displacement dependence at the upper spinodal. The second-order and spinodal duration data are consistent with these local estimates over the resolved ranges; at larger displacements the local expansion degrades, as expected from the higher-order potential terms. It is important to note that while the scaling laws of the duration are robust, the absolute normalization of \(\Delta t\) explicitly depends on the chosen linearity threshold. Therefore, these interval durations should be interpreted as threshold-dependent measures of potential flatness, rather than definition-independent equilibration times. These local analytical predictions are fully cross-validated by the full 5D numerical evolution, which confirms the genericity of the slow-roll dynamics without encountering numerical instabilities at the horizon.

In a broader context, our findings establish a direct geometric bridge to the prethermalization dynamics observed in 4D quenched field theories~\cite{Chiocchetta2015, Chiocchetta2017}. In both cases, a sudden quench generates a long-lived intermediate stage that loses memory of the initial conditions. However, the underlying protective mechanisms are fundamentally different. In 4D weakly coupled systems, the prethermal state is governed by dynamical renormalization group fixed points, and the order parameter exhibits power-law or logarithmic time dependence determined by the associated critical exponents. In the strongly coupled holographic plasma considered here, the protective role is instead played by the local geometric flatness of the phase-transition endpoints. The ``double-zero'' condition (\(V'=0\), \(V''=0\)) flattens the effective potential, causing both the drift velocity and its acceleration to vanish. This dual perspective---RG fixed points versus geometric flatness---enriches the general understanding of quench dynamics across different coupling regimes.

A notable phenomenological consequence of this structurally protected stage is its extended duration, which can reach up to \(10^4/(\pi T)\). Because the chiral condensate remains out of equilibrium for such a parametrically long time, the effective masses of the quarks are continuously modulated during the medium's evolution. This prolonged slow roll can naturally leave distinctive imprints on experimental observables. In particular, electromagnetic probes such as soft photons and dileptons are highly sensitive to the time-dependent quark effective mass and the non-equilibrium spectral functions of the medium; the persistent drifting of the condensate may lead to a parametric enhancement of low-invariant-mass dilepton yields relative to equilibrium expectations. Moreover, the structural distinction in the drift laws implies that the out-of-equilibrium evolution near different phase-transition endpoints follows distinctly different trajectories. This extended stage also provides a wide time window for critical fluctuations to develop and persist, which could affect the higher-order moments of final-state hadron distributions in heavy-ion collisions.

Finally, we emphasize the current limitations of our study and outline promising directions for future research. The present calculation is restricted to a fixed background with a probe scalar, a homogeneous condensate, and a constant kinetic coefficient. Spatial fluctuations, nucleation, conserved-density dynamics, and backreaction~\cite{Randrup2010Spinodal,Janik2017Separation,Attems2020PhaseSeparation,Zhao2024Spinodal,Motta2025Inhomogeneous} are essential to fully understand spinodal decomposition and inhomogeneous chiral instabilities. In particular, extending this work to spatially inhomogeneous modes naturally connects to the physics of the disoriented chiral condensate~\cite{Rajagopal:1992qz,Rajagopal:1993ah}. In a quench scenario, the ultra-long lifetime of the slow roll discovered here provides an exceptionally wide time window for the parametric amplification of long-wavelength pion modes. It would be highly interesting to investigate whether the ``double-zero'' background significantly alters the growth rate or the resulting domain size of the DCC. Furthermore, extending the current analysis to a finite quark mass would break the strict \(Z_2\) symmetry, turning the tricritical point into a critical end point (CEP) or a crossover. On the methodological side, computing the non-equilibrium spectral functions on top of this long-lived slow-roll background and exploring their interplay with chiral anomalous transport (such as the chiral magnetic effect) will provide a more complete phenomenological picture. Connecting the holographic dynamical evolution with experimentally measured soft photon and dilepton spectra will ultimately bridge the gap between the geometric structure of the phase transition and the observable signals in strongly coupled QCD.

\vspace*{1cm}

{\bf Acknowledgements}\quad
This work is mainly supported by the National Natural Science Foundation of China under Grant Nos. 12275108, 12235016.

\appendix

\section{Constrained determination of the coefficients}\label{app:finite_temperature_reduction}

This appendix determines the reduced drift coefficients from the bulk evolution equation with vanishing boundary source and future-horizon regularity. At fixed temperature and chemical potential, the slow field profile and its drift satisfy
\begin{equation}
\label{eq:appendix_invariance_equation}
\mathcal F\,\mathcal D[\partial_\sigma\bar\chi]=R[\bar\chi],
\qquad \mathcal F=\dot\sigma,
\end{equation}
where
\begin{equation}
\label{eq:appendix_bulk_operators}
\begin{aligned}
\mathcal D&=2\partial_z-\left(\frac3z+\Phi'\right),\\
R[h]&=f h''-\left(\frac{3f}{z}+f\Phi'-f'\right)h'\\
&\quad-\frac{2v_2(z)h+4v_4(z)h^3+6v_6(z)h^5}{z^2}.
\end{aligned}
\end{equation}
Primes denote derivatives with respect to $z$. The functions $f$, $\Phi$, and $v_i$ are evaluated on the chosen background. At $\mu=0$, $v_2=-3/2$, $v_4=v_{40}$, and $v_6=v_{60}$. Their radial dependence is retained at finite chemical potential. We define the physical-condensate functional by
\begin{equation}
\mathcal B[h]=\zeta\lim_{z\to0}\frac{h(z)}{z^3}.
\end{equation}

\subsection{Expansion about the zero-condensate background}

Near the second-order critical point and the lower spinodal, we expand about $\chi=0$. Chiral symmetry permits odd powers of the physical condensate. Near the second-order transition, the retained terms are
\begin{equation}
\label{eq:appendix_second_order_expansion}
\begin{aligned}
\bar\chi(z,\sigma;T)&=\sigma\chi_1(z,T)+\sigma^3\chi_3(z,T)+O(\sigma^5),\\
\mathcal F(\sigma,T)&=c_1(T)\sigma+c_3(T)\sigma^3+O(\sigma^5).
\end{aligned}
\end{equation}
Near the lower spinodal, we retain the fifth-order terms,
\begin{equation}
\label{eq:appendix_lower_expansion}
\begin{aligned}
\bar\chi(z,\sigma;T)&=\sigma\chi_1(z,T)+\sigma^3\chi_3(z,T)\\
&\quad+\sigma^5\chi_5(z,T)+O(\sigma^7),\\
\mathcal F(\sigma,T)&=c_1(T)\sigma+c_3(T)\sigma^3\\
&\quad+c_5(T)\sigma^5+O(\sigma^7).
\end{aligned}
\end{equation}
The coefficients and radial functions are determined separately on each background. At $T_c$ and $T_{\mathrm{sp}}^-$, $c_1=0$, while $c_1(T)$ is retained at nearby temperatures. The two expansions use the same constrained construction through cubic order. The fifth-order problem is additionally required for the lower-spinodal calculation. Temperature arguments are suppressed below.

For an expansion about the zero-condensate background, the radial coefficients satisfy
\begin{equation}
\label{eq:appendix_boundary_normalization}
\begin{aligned}
\lim_{z\to0}\frac{\chi_j(z)}{z}&=0,
\qquad j=1,3,5,\\
\mathcal B[\chi_1]&=1,
\qquad
\mathcal B[\chi_3]=\mathcal B[\chi_5]=0.
\end{aligned}
\end{equation}
The first condition removes the boundary source. The remaining conditions fix the leading amplitude and ensure that higher-order radial corrections do not change the physical condensate. All radial functions are also required to be smooth at the future horizon in ingoing coordinates. 

Substitution into Eq.~\eqref{eq:appendix_invariance_equation} at linear order gives the generalized eigenvalue problem
\begin{equation}
\label{eq:appendix_linear_eigenproblem}
\mathcal L\chi_1=c_1\mathcal D\chi_1.
\end{equation}
The linearized radial operator is
\begin{equation}
\begin{aligned}
\mathcal L
&=f\partial_z^2
-\left(\frac{3f}{z}+f\Phi'-f'\right)\partial_z
-\frac{2v_2(z)}{z^2},
\end{aligned}
\end{equation}
and horizon regularity imposes
\begin{equation}
\label{eq:appendix_linear_horizon}
\left.
\left\{
(f'-2c_1)\chi_1'
+\left[
c_1\left(\frac3z+\Phi'\right)-\frac{2v_2}{z^2}
\right]\chi_1
\right\}
\right|_{z=z_h}=0.
\end{equation}
The source-free boundary condition and horizon regularity select the allowed eigenvalues. We select the branch connected to the marginal mode at the second-order critical point or the lower spinodal. Its eigenvalue is $c_1(T)$, and $\mathcal B[\chi_1]=1$ fixes the amplitude of the corresponding eigenfunction.

Once $c_1$ and $\chi_1$ are determined, the cubic invariance equation gives
\begin{equation}
\label{eq:appendix_cubic_constrained}
(\mathcal L-3c_1\mathcal D)\chi_3
-c_3\mathcal D\chi_1
=\frac{4v_4(z)}{z^2}\chi_1^3.
\end{equation}
The unknown function $\chi_3$ and constant $c_3$ are determined simultaneously using Eq.~\eqref{eq:appendix_boundary_normalization} and
\begin{equation}
\label{eq:appendix_cubic_horizon}
\begin{aligned}
\Bigg[
&(f'-6c_1)\chi_3'
+\left[
3c_1\left(\frac3z+\Phi'\right)-\frac{2v_2}{z^2}
\right]\chi_3\\
&-c_3\mathcal D\chi_1
-\frac{4v_4}{z^2}\chi_1^3
\Bigg]_{z=z_h}=0.
\end{aligned}
\end{equation}
When $c_1=0$, let $\eta_0$ satisfy $\mathcal L^\dagger\eta_0=0$, with adjoint boundary conditions that eliminate the boundary terms. Using the radial pairing $\langle a,b\rangle=\int_0^{z_h}a^*(z)b(z)\,dz$, projection of Eq.~\eqref{eq:appendix_cubic_constrained} gives
\begin{equation}
\label{eq:appendix_cubic_projection_general}
c_3=-4\frac{\langle\eta_0,z^{-2}v_4(z)\chi_1^3\rangle}
{\langle\eta_0,\mathcal D\chi_1\rangle}.
\end{equation}
At $T_c$ and $\mu=0$, this becomes
\begin{equation}
\label{eq:critical_cubic_coefficient}
c_3=-4v_{40}
\frac{\langle\eta_0,z^{-2}\chi_1^3\rangle}
{\langle\eta_0,\mathcal D\chi_1\rangle}.
\end{equation}
At nonzero $c_1$, the term $3c_1\mathcal D\chi_3$ is retained in the constrained problem.

The fifth-order invariance equation similarly yields
\begin{equation}
\label{eq:appendix_quintic_constrained}
\begin{aligned}
(\mathcal L-5c_1\mathcal D)\chi_5
-c_5\mathcal D\chi_1
={}&3c_3\mathcal D\chi_3\\
&+\frac{12v_4(z)}{z^2}\chi_1^2\chi_3
+\frac{6v_6(z)}{z^2}\chi_1^5.
\end{aligned}
\end{equation}
Its right-hand side is fixed by the preceding calculations. The boundary conditions in Eq.~\eqref{eq:appendix_boundary_normalization} and the horizon relation
\begin{equation}
\label{eq:appendix_quintic_horizon}
\begin{aligned}
\Bigg[
&(f'-10c_1)\chi_5'
+\left[
5c_1\left(\frac3z+\Phi'\right)-\frac{2v_2}{z^2}
\right]\chi_5\\
&-c_5\mathcal D\chi_1
-3c_3\mathcal D\chi_3
-\frac{12v_4}{z^2}\chi_1^2\chi_3
-\frac{6v_6}{z^2}\chi_1^5
\Bigg]_{z=z_h}=0
\end{aligned}
\end{equation}
determine $\chi_5$ and $c_5$. Each higher-order problem is linear in the radial correction and its associated drift coefficient and can be solved as an augmented linear system after radial discretization.

At $c_1=0$, projection of Eq.~\eqref{eq:appendix_quintic_constrained} onto the corresponding adjoint zero mode gives
\begin{equation}
\label{eq:quintic_projection_coefficient}
\begin{aligned}
c_5={}&-\frac{1}{\langle\eta_0,\mathcal D\chi_1\rangle}
\Bigl[3c_3\langle\eta_0,\mathcal D\chi_3\rangle\\
&+12\langle\eta_0,z^{-2}v_4(z)\chi_1^2\chi_3\rangle\\
&+6\langle\eta_0,z^{-2}v_6(z)\chi_1^5\rangle\Bigr].
\end{aligned}
\end{equation}
Here $\chi_3$ is obtained from the cubic constrained problem with $\mathcal B[\chi_3]=0$. At the lower spinodal, the zero mode and all radial functions refer to that endpoint background. For $c_1\ne0$, the full constrained equations are used instead of these zero-mode projection formulas.

\subsection{Expansion about the upper-spinodal background}

At the upper spinodal, we expand about a nonzero background. The reference profile is the nonzero stationary solution $\chi_{\mathrm{sp}}^+$ satisfying $R_{\mathrm{sp}}^+[\chi_{\mathrm{sp}}^+]=0$, where $R_{\mathrm{sp}}^+$ denotes $R$ evaluated at $T_{\mathrm{sp}}^+$. Its source-free boundary behavior is $\chi_{\mathrm{sp}}^+(z)\sim\sigma_{\mathrm{sp}}^+z^3/\zeta$, with $\sigma_{\mathrm{sp}}^+=\sigma_{\mathrm{eq}}(T_{\mathrm{sp}}^+)$. We use
\begin{equation}
\delta\sigma=\sigma-\sigma_{\mathrm{sp}}^+,
\qquad
\epsilon=\frac{T-T_{\mathrm{sp}}^+}{T_{\mathrm{sp}}^+},
\end{equation}
and expand
\begin{equation}
\begin{aligned}
\bar{\chi}
={}&\chi_{\mathrm{sp}}^+
+\delta\sigma\,\chi_1
+\epsilon\chi_\epsilon
+(\delta\sigma)^2\chi_2+\cdots,\\
\mathcal F(\delta\sigma,\epsilon)
={}&a\epsilon+b(\delta\sigma)^2
+O\!\left(\epsilon\delta\sigma,(\delta\sigma)^3,\epsilon^2\right).
\end{aligned}
\end{equation}
At fixed temperature, $\mathcal F=\dot\sigma=\dot{\delta\sigma}$. The global chiral symmetry relates the positive and negative endpoint backgrounds but does not require the drift about a single nonzero endpoint to be odd in $\delta\sigma$. In the remainder of this subsection, $\chi_1$ denotes the tangent mode at the upper spinodal, and all background functions are evaluated at $T_{\mathrm{sp}}^+$. The endpoint operators are
\begin{equation}
\begin{aligned}
\mathcal D_{\mathrm{sp}}^+
={}&2\partial_z-\left(\frac3z+\Phi'\right),\\
\mathcal L_{\mathrm{sp}}^+
={}&f\partial_z^2
-\left(\frac{3f}{z}+f\Phi'-f'\right)\partial_z\\
&-\frac{
2v_2+12v_4(\chi_{\mathrm{sp}}^+)^2
+30v_6(\chi_{\mathrm{sp}}^+)^4
}{z^2}.
\end{aligned}
\end{equation}

The tangent mode satisfies
\begin{equation}
\mathcal L_{\mathrm{sp}}^+\chi_1=0.
\end{equation}
The source and normalization conditions are
\begin{equation}
\label{eq:appendix_upper_boundary}
\begin{aligned}
\lim_{z\to0}\frac{\chi_1}{z}
=\lim_{z\to0}\frac{\chi_\epsilon}{z}
=\lim_{z\to0}\frac{\chi_2}{z}&=0,\\
\mathcal B[\chi_1]=1,
\qquad
\mathcal B[\chi_\epsilon]=\mathcal B[\chi_2]&=0.
\end{aligned}
\end{equation}
For $\chi_1$, horizon regularity gives
\begin{equation}
\label{eq:appendix_upper_mode_horizon}
\left[
f'\chi_1'
-\frac{
2v_2+12v_4(\chi_{\mathrm{sp}}^+)^2
+30v_6(\chi_{\mathrm{sp}}^+)^4
}{z^2}\chi_1
\right]_{z=z_h}=0.
\end{equation}

Substitution into Eq.~\eqref{eq:appendix_invariance_equation} gives the temperature-order problem
\begin{equation}
\label{eq:appendix_upper_temperature}
\begin{aligned}
\mathcal L_{\mathrm{sp}}^+\chi_\epsilon
-a\mathcal D_{\mathrm{sp}}^+\chi_1
=-\left.
\partial_\epsilon
R_{T_{\mathrm{sp}}^+(1+\epsilon)}
[\chi_{\mathrm{sp}}^+]
\right|_{\epsilon=0}.
\end{aligned}
\end{equation}
The temperature derivative is taken at fixed endpoint profile after mapping neighboring radial domains to a common interval while preserving the physical-condensate normalization. The same convention is used when evaluating the derivative at the horizon. Together with Eq.~\eqref{eq:appendix_upper_boundary}, the regularity condition is
\begin{equation}
\label{eq:appendix_upper_temperature_horizon}
\begin{aligned}
\Bigg[
&f'\chi_\epsilon'
-\frac{
2v_2+12v_4(\chi_{\mathrm{sp}}^+)^2
+30v_6(\chi_{\mathrm{sp}}^+)^4
}{z^2}\chi_\epsilon\\
&-a\mathcal D_{\mathrm{sp}}^+\chi_1
+\left.
\partial_\epsilon
R_{T_{\mathrm{sp}}^+(1+\epsilon)}
[\chi_{\mathrm{sp}}^+]
\right|_{\epsilon=0}
\Bigg]_{z=z_h}=0.
\end{aligned}
\end{equation}
These conditions determine $a$ and $\chi_\epsilon$.

At quadratic order in $\delta\sigma$, Eq.~\eqref{eq:appendix_invariance_equation} gives
\begin{equation}
\label{eq:appendix_upper_quadratic}
\mathcal L_{\mathrm{sp}}^+\chi_2
-b\mathcal D_{\mathrm{sp}}^+\chi_1
=
\frac{
12v_4\chi_{\mathrm{sp}}^+
+60v_6(\chi_{\mathrm{sp}}^+)^3
}{z^2}\chi_1^2.
\end{equation}
The corresponding horizon condition is
\begin{equation}
\label{eq:appendix_upper_quadratic_horizon}
\begin{aligned}
\Bigg[
&f'\chi_2'
-\frac{
2v_2+12v_4(\chi_{\mathrm{sp}}^+)^2
+30v_6(\chi_{\mathrm{sp}}^+)^4
}{z^2}\chi_2\\
&-b\mathcal D_{\mathrm{sp}}^+\chi_1
-\frac{
12v_4(\chi_{\mathrm{sp}}^+)
+60v_6(\chi_{\mathrm{sp}}^+)^3
}{z^2}\chi_1^2
\Bigg]_{z=z_h}=0.
\end{aligned}
\end{equation}
Together with Eq.~\eqref{eq:appendix_upper_boundary}, this determines $b$ and $\chi_2$. Let $\eta_{\mathrm{sp}}^+$ satisfy $(\mathcal L_{\mathrm{sp}}^+)^{\dagger}\eta_{\mathrm{sp}}^+=0$, with adjoint boundary conditions that eliminate the boundary terms. Using the same radial pairing, projection of Eqs.~\eqref{eq:appendix_upper_temperature} and~\eqref{eq:appendix_upper_quadratic} yields
\begin{equation}
\label{eq:upper_spinodal_coefficients}
\begin{aligned}
a&=\frac{
\left\langle\eta_{\mathrm{sp}}^+,
\left.\partial_\epsilon R_{T_{\mathrm{sp}}^+(1+\epsilon)}
[\chi_{\mathrm{sp}}^+]\right|_{\epsilon=0}
\right\rangle}
{\langle\eta_{\mathrm{sp}}^+,\mathcal D_{\mathrm{sp}}^+\chi_1\rangle},\\
b&=-\frac{
\left\langle\eta_{\mathrm{sp}}^+,
z^{-2}\left(12v_4\chi_{\mathrm{sp}}^+
+60v_6(\chi_{\mathrm{sp}}^+)^3\right)\chi_1^2\right\rangle}
{\langle\eta_{\mathrm{sp}}^+,\mathcal D_{\mathrm{sp}}^+\chi_1\rangle}.
\end{aligned}
\end{equation}
The coefficients $a$ and $b$ are evaluated at $T_{\mathrm{sp}}^+$ and retained as constants in the leading expansion in $\epsilon$ and $\delta\sigma$. All reduced coefficients are obtained from the bulk equations and boundary conditions independently of the real-time trajectories used in the subsequent comparisons.

\section{Slow-roll duration from the time expansion}
\label{app:duration_derivation}

This appendix derives the analytic criterion for the slow-roll duration \(\Delta t\) based on the short-time expansion of the effective potential. At fixed temperature, Eq.~\eqref{eq:GL_with_mobility} gives the expansion about the entrance time $t_L$. With $\tau=t-t_L$, retaining terms through $\tau^3$ yields
\begin{equation}
\label{eq:duration_cubic_time_expansion}
\begin{split}
\sigma(t_L+\tau)={}&\sigma_L-U_{\mathrm{eff}}'\tau
+\frac12 U_{\mathrm{eff}}'U_{\mathrm{eff}}''\tau^2\\
&-\frac16 U_{\mathrm{eff}}'\left[(U_{\mathrm{eff}}'')^2
+U_{\mathrm{eff}}'U_{\mathrm{eff}}'''\right]\tau^3
+O(\tau^4).
\end{split}
\end{equation}
All potential derivatives in this appendix are evaluated at $(\sigma_L,T)$. The goodness of this linear fit is quantified by the standard coefficient of determination, \(R^2 = 1 - SS_{\mathrm{res}}/SS_{\mathrm{tot}}\), where \(SS_{\mathrm{res}}\) is the residual sum of squares and \(SS_{\mathrm{tot}}\) is the total variance of the trajectory. In the continuum limit of uniform time sampling, direct integration yields the exact analytic expression:
\begin{equation}
\label{eq:duration_cubic_r_squared}
\begin{aligned}
R^2(\Delta t)
={}&\Biggl\{
1-\frac12 U_{\mathrm{eff}}''\Delta t\\
&+\frac{3}{20}
\left[
(U_{\mathrm{eff}}'')^2
+U_{\mathrm{eff}}'U_{\mathrm{eff}}'''
\right](\Delta t)^2
\Biggr\}^{2}\\
&\times\Biggl\{
1-U_{\mathrm{eff}}''\Delta t\\
&+\frac{17}{30}(U_{\mathrm{eff}}'')^2(\Delta t)^2
+\frac{3}{10}U_{\mathrm{eff}}'U_{\mathrm{eff}}'''(\Delta t)^2\\
&-\frac16 U_{\mathrm{eff}}''
\left[
(U_{\mathrm{eff}}'')^2
+U_{\mathrm{eff}}'U_{\mathrm{eff}}'''
\right](\Delta t)^3\\
&+\frac{3}{112}
\left[
(U_{\mathrm{eff}}'')^2
+U_{\mathrm{eff}}'U_{\mathrm{eff}}'''
\right]^2
(\Delta t)^4
\Biggr\}^{-1}.
\end{aligned}
\end{equation}
For a given entrance time $t_L$, we extend the fitting interval toward later times. The estimated endpoint $t_R$ is determined by the first downward crossing of the linearity threshold, giving
\begin{equation}
\label{eq:duration_cubic_threshold}
R^2(t_R-t_L)=R_{\mathrm{th}}^2,
\qquad R_{\mathrm{th}}^2=0.99999.
\end{equation}
The corresponding duration is $\Delta t=t_R-t_L$. Here, $R^2$ is evaluated from Eq.~\eqref{eq:duration_cubic_r_squared}.

When the quadratic correction dominates, the leading departure from linearity is
\begin{equation}
\label{eq:duration_quadratic_limit}
1-R^2\simeq\frac{(U_{\mathrm{eff}}'')^2(\Delta t)^2}{60},
\end{equation}
At a nonstationary inflection point with $U_{\mathrm{eff}}''=0$ and $U_{\mathrm{eff}}'U_{\mathrm{eff}}'''\ne0$, the cubic correction instead gives
\begin{equation}
\label{eq:duration_inflection_limit}
\Delta t\simeq
\left[\frac{700(1-R_{\mathrm{th}}^2)}{3}\right]^{1/4}
\frac{1}{\sqrt{|U_{\mathrm{eff}}'U_{\mathrm{eff}}'''|}}.
\end{equation}
Thus, the third-order time term sets a finite duration when the local curvature vanishes but the condensate continues to evolve.

\bibliographystyle{apsrev4-1}
\bibliography{refs}
\end{document}